\documentclass[manuscript]{emulateapj}
\usepackage{natbib}
\usepackage{color}
\usepackage{amsmath}
\usepackage{array}
\newcolumntype{P}[1]{>{\centering\arraybackslash}p{#1}}

\usepackage[toc,page]{appendix}
\usepackage{afterpage}
\usepackage{float}
\usepackage{capt-of}
\usepackage{bm}
\usepackage[dvipsnames]{xcolor}
\usepackage{esint}
\usepackage{ mathrsfs }

\shortauthors{Carr, Michel-Dansac, Blaizot, Scarlata, Henry, \& Verhamme}

\begin{document}

\title{Testing SALT Approximations with Numerical Radiation Transfer Code Part 1: Validity and Applicability}

\author{C. Carr\altaffilmark{1}, L. Michel-Dansac\altaffilmark{2}, J. Blaizot\altaffilmark{2}, C. Scarlata\altaffilmark{1}, A. Henry\altaffilmark{4}, A. Verhamme\altaffilmark{3,2}}

 \altaffiltext{1}{Minnesota Institute for Astrophysics, School of
   Physics and Astronomy, University of Minnesota, 316 Church str
 SE, Minneapolis, MN 55455,USA}
\altaffiltext{2}{Univ Lyon, Univ Lyon1, ENS de Lyon, CNRS, Centre de Recherche Astrophysique de Lyon UMR5574, 69230 Saint-Genis-Laval,
France}
\altaffiltext{3}{Observatoire de Genève, Université de Genève, 51 Ch. des Maillettes, 1290 Versoix, Switzerland}
\altaffiltext{4}{Space Telescope Science Institute, 3700 San Martin
   Drive, Baltimore, MD 21218, USA,  panagia@stsci.edu}

\begin{abstract}
Absorption line spectroscopy offers one of the best opportunities to constrain the properties of galactic outflows and the environment of the circumgalactic medium.  Extracting physical information from line profiles is difficult, however, for the physics governing the underlying radiation transfer is complicated and depends on many different parameters.  Idealised analytical models are necessary to constrain the large parameter spaces efficiently, but are typically plagued by model degeneracy and systematic errors.  Comparison tests with idealized numerical radiation transfer codes offer an excellent opportunity to confront both of these issues.  In this paper, we present a detailed comparison between SALT, an analytical radiation transfer model for predicting UV spectra of galactic outflows, with the numerical radiation transfer software, RASCAS.  Our analysis has lead to upgrades to both models including an improved derivation of SALT and a customizable adaptive mesh refinement routine for RASCAS.  We explore how well SALT, when paired with a Monte Carlo fitting procedure, can recover flow parameters from non-turbulent and turbulent flows.  When the velocity and density gradients are excluded, we find that flow parameters are well recovered from high resolution ($20 \ \rm{km} \ \rm{s}^{-1}$) data and moderately well from medium resolution ($100 \ \rm{km} \ \rm{s}^{-1}$) data without turbulence at a $\rm{S/N} = 10$, while derived quantities (e.g., mass outflow rates, column density, etc.) are well recovered at all resolutions.  In the turbulent case, biased errors emerge in the recovery of individual parameters, but derived quantities are still well recovered.  

\end{abstract}

\section{Introduction}

Spectroscopic studies over the past decade have made remarkable progress towards revealing the properties of the circumgalactic medium (CGM), or the gaseous halo surrounding a galaxy, separating the interstellar medium (ISM) from the intergalactic medium (IGM).  Both space and ground based instruments have provided critical observations in the optical \citep{Martin2012,Rubin2014,Henry2018,Eggen2021}, sub mm \citep{Cicone2015,Gallerani2018}, UV \citep{Werk2013,Zhu2015,Heckman2015,Henry2015,Prochaska2017,Berg2022,Saldana-Lobez2022,Xu2022}, and x-ray \citep{Laha2018} revealing the CGM to be multi-phase, dynamic, and metal enriched.  The current paradigm posits that massive multi-phase flows - spanning up to six orders of magnitude in temperature ($10-10^{7} \ \rm{K}$) - are constantly moving pristine and metal enriched gas in and out of the CGM at all redshifts \citep{Somerville2015,Rupke2018,Veilleux2020}.  This transport of material by flows is fundamental to the baryon cycle and surrounding ecosystems.  Indeed, the metals entrained in flows can radically alter the properties of a plasma by changing the cooling time and hence the dynamics of the gas.  

While we have drawn back the curtain to reveal the primary features of the CGM, precise measurements of the properties of flows are still lacking, and are a top priority for astronomers in the coming decade (see priority area: unveiling the drivers of galaxy growth, \citealt{Astro2022}).  Future instruments such as the 30 meter class telescopes and the 6 meter class UV space telescope will greatly improve on our current observational limitations in terms of resolution and collecting area, giving astronomers the necessary boost in performance to place tighter constraints on the properties of flows.

The modeling of absorption and emission lines in down-the-barrel spectroscopy (i.e., centered on the object of study) represents one of the most promising approaches to constraining the properties of flows.  The robustness and scope of predictions, however, depends on the complexity of the model used (or lack thereof).  On one end, empirical based measurements such as EWs, FWHMs, centroid velocities, etc., can be used to successfully establish the presence of flows by placing constraints on basic kinematic properties such as terminal speeds, line widths, etc., (e.g., \citealt{Heckman2015,Henry2015,Chisholm2016a}); while on the other end, models of the underlying radiation transfer (e.g., \citealt{Chisholm2015,Scarlata2015,Zhu2015,Krumholz2017,Carr2018}, 2022 in prep) have the potential to constrain a much larger range of phenomena including outflow rates \citep{Chisholm2016b,Chisholm2017b,Xu2022}, different driving mechanisms \citep{Yuan2022}, and escape fractions of ionizing radiation \citep{Gazagnes2018,Chisholm2018,Chisholm2020}.  The latter of which have the potential to constrain the Epoch of Reionizaiton.  

The art or challenge behind this process is to design a model which is simple enough to constrain its parameter space quickly while still maintaining enough complexity to capture the underlying physics.  Typically, as models grow in complexity, degeneracy appears, and higher quality data is often necessary (if even possible, see \citealt{Gronke2015}) to distinguish between different model predictions.  

To achieve analytical simplicity, the radiation transfer models used to interpret spectral lines typically rely on analytical approximations and idealized model configurations to solve the radiation transfer equation.  This contrasts with numerical methods, which often treat the radiation transfer process in full.  While numerical codes are computationally expensive, they do provide an excellent opportunity to test the simpler analytical models (e.g., \citealt{Gronke2016}) with potential benefits to both participants.  In this paper, we test the semi-analytical line transfer (SALT) model of \cite{Scarlata2015} against the numerical radiation transfer code, RASCAS.  SALT (and similar models) have been and will be used in future studies to estimate mass outflow rates (e.g., Huberty et al. 2023 in prep), constrain the neutral gas distribution to infer the LyC escape fraction (e.g., \citealt{Carr2021HST}), and to understand the relation between metal line profiles and Ly$\alpha$ scattering (e.g., \citealt{Carr2021}, 2023 in prep). For these reasons, it is important that SALT is tested to the best of our ability.

Inspired by our comparison, we have developed upgrades to both models, and present them here for the first time.  These include an alternative derivation of the SALT model which accounts for the absorption of non-radially traveling photons by an expanding shell and a customizable adaptive mesh refinement (AMR) routine for RASCAS.  Because RASCAS is a mesh based radiation transfer code, it depends on the natural resolution associated with the dimensions of the mesh.   Therefore, using the SALT model as a guide, we develop an AMR routine capable of resolving features of idealized flows in RASCAS.

SALT relies on the Sobolev approximation \citep{Lamers1999,Scarlata2015}, and does not account for turbulent/thermal line broadening when computing line profiles.  As such, turbulence is a key physical aspect of reality missing from the SALT model.  RASCAS, on the other hand, does account for turbulent/thermal line broadening when computing line profiles thereby creating the perfect opportunity to establish SALT's range of validity and applicability.  While the limitations of the Sobolev approximation in the context of line profiles is well known (e.g., \citealt{deKoter1993,Owocki1999}), it has not been studied in the SALT formalism before.  Therefore, we review the formalism behind the Sobolev approximation and establish a validity criterion for SALT.  We then test how well SALT can recover the bulk properties of turbulent flows from mock spectra generated with RASCAS.

The rest of this paper is organized as follows.  We present an improved derivation of the SALT model in section 2 and introduce the RASCAS customizable AMR routine in section 3.  We then compare spectral line predictions of the upgraded versions of SALT and RASCAS in the limit of low turbulent/thermal motion in section 4.  In section 5, we review the formalism behind the Sobolev approximation and establish a validity criterion in terms of the SALT parameter space.  We then test how well SALT can recover the bulk properties of turbulent flows in section 6.  Finally, we give our conclusions in section 7.

\section{SALT Upgrade: Non-radial Absorption}
\label{Formalism}

\begin{figure}
  \centering
\includegraphics[scale=.4]{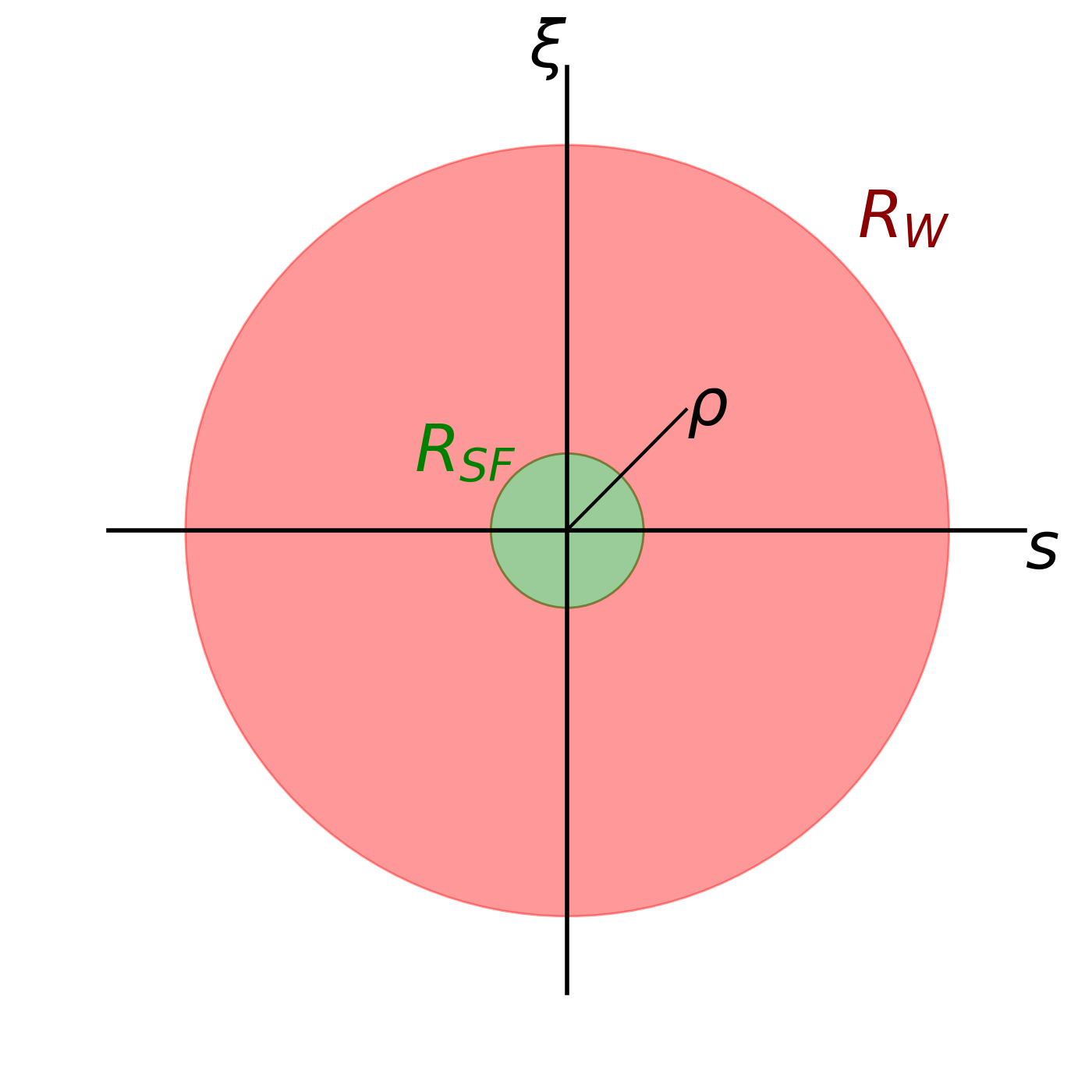}
 \caption{Envelope of material (shown in red) of radius $R_W$ surrounding a galaxy (shown in green) of radius $R_{SF}$.  The $\xi$ and $s$-axes are written in normalized units $(r/R_{SF})$.  $\rho$ represents an arbitrary radius written in normalized units.  The envelope represents an outflow characterized by a density and velocity field.}
   \label{f1}
\end{figure} 

\begin{figure*}
  \centering
\includegraphics[scale=.8]{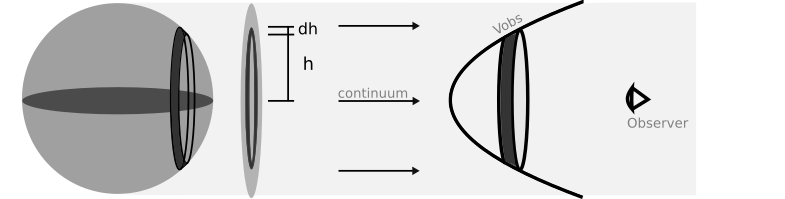}
 \caption{A spherical source of isotropically emitted radiation surrounded by an outflow of material.  The continuum consists only of rays emitted parallel to the line of sight.  We assume the continuum radiation is projected uniformly onto the plane of the sky, and approximate the source as a uniformly emitting disk.  The structure on the far right represents a surface of constant observed velocity, $\Omega_v$, for a velocity field of $\gamma = 0.5$.  Radiation passing through this surface at its resonant wavelength will be removed from the continuum.}
   \label{f2}
\end{figure*} 

To achieve analytical simplicity, interpretive radiation transfer models must rely on approximations to solve the radiation transfer equation, accommodate for the scattering of photons, etc.  The ability to test these approximations against numerical code is powerful with benefits for each participant.  Motivated by our desire to better match the results of numerical scattering experiments, in this section we present an alternative derivation of the SALT model originally presented by \cite{Scarlata2015}.  Physically, the new model accounts for the absorption of photons traveling in non-radial directions by the outflow.  Since these photons were neglected in earlier versions of SALT, this new model represents an upgrade to the older versions.  In addition, we relax the assumption of a constant mass outflow rate (e.g., \citealt{Carr2018,Carr2021}) by accounting for a density field of arbitrary power law.   

For consistency, we preserve what notation we can from previous works.  All models are to assume a spherical source of isotropically emitted radiation of radius, $R_{SF}$, surrounded by an envelope of outflowing material which extends to a terminal radius, $R_W$.  A diagram has been provided in Figure~\ref{f1}.  The $\xi$-axis runs perpendicular to the line of sight and is measured using normalized units, $r/RSF$.  The $s$-axis runs parallel with the line of sight and is measured using the same normalized units.  We refer to an arbitrary radius measured in the normalized units as $\rho$.  

\subsection{New Absorption Profile}
In the original version of the SALT model, the outflow is decomposed into shells and the observed spectrum is computed one shell at a time.  In the case of the absorption profile, the total energy removed from the continuum by a single shell is approximated by considering only radially emitted photons absorbed at resonance by the shell.  In other words, it was assumed that a single shell can only absorb photons at a single wavelength.  In reality, however, for an extended source, a single shell is capable of absorbing non-radially traveling photons from a range of wavelengths.  Thus, to improve upon the original SALT model, we seek to account for the additional absorption of these photons.  To this end, we abandon the approach of decomposing the outflow into thin shells, and instead, focus on calculating the total amount of energy removed from the continuum at a single wavelength from all points of resonance in the outflow.      

Since the observer is located very far from the source, we can assume that the continuum consists only of photons emitted parallel to the line of sight.  For simplicity, we also assume that the source projects continuum photons uniformly onto the plane of the sky - that is, we treat the source as a uniform emitting area in the shape of a circle.  A diagram has been provided in Figure~\ref{f2}.  We assume the outflow is characterized according to the description of \cite{Carr2018}.  In particular, we assume a velocity field of power law, 
\begin{equation}
\begin{aligned}
v &= v_0\left(\frac{r}{R_{\text{SF}}}\right)^{\gamma} &&\text{for}\ r < R_{W} \\[1em]
v &= v_{\infty}  &&\text{for} \ r \geq R_{W}, \\
\end{aligned}
\label{Velocity_Equation}
\end{equation}  
\noindent where $v_0$ is the wind velocity at $R_{SF}$, $v_{\infty}$ is the terminal velocity of the wind, and $0.5<\gamma<2$ is a typical range for an accelerating velocity field (e.g., \citealt{Scuderi1992,Carr2018,Carr2021}).  Lastly, we adopt the Sobolev approximation \citep{Lamers1999}, and assume only photons propagating exactly at resonance with a given point in the outflow can be absorbed at that point.

Let photons of wavelength, $\lambda$, be at resonance with material moving at velocity, $v$.  Then only points in the outflow with observed velocity, $v_{obs}$ (i.e., the value of the velocity field projected onto the line of sight), equal to $v$, can absorb photons from the continuum at wavelength, $\lambda$.  Geometrically, these points form a smooth surface.  We refer to this surface, $ \Omega_v$, as the surface of constant observed velocity, $v$.  The intersection, $\Gamma_v$, of such a surface with the $s \xi$-plane is shown to the right in Figure~\ref{f2}.  

Our goal now is to calculate the fraction of the continuum energy emitted at wavelength, $\lambda$, and absorbed by $\Omega_v$.  To achieve this goal, we consider the intensity emitted from a ring centered on the $s$-axis with radius, $h$.  To compute the normalized spectrum in the absence of the outflow, we would have
\begin{eqnarray}
I(h)_{\rm{abs,blue}}/I_0 &=& \frac{F_{\lambda}}{F_{c,\lambda}}\frac{\int_0^{1}2\pi hdh}{\pi (1)^2} \\
&=&\frac{F_{\lambda}}{F_{c,\lambda}},
\end{eqnarray} 
where $F_{\lambda}/F_{c,\lambda}$ is the flux observed at wavelength, $\lambda$, normalized by the continuum\footnote{For a flat continuum, $F_{\lambda}/F_{c,\lambda} =1 $, however, this may not hold for all line profiles of interest - for example, in the cases of overlapping absorption profiles, nebular lines, absorption in the ISM, etc.}.  The fraction of energy removed from each ring by $\Omega_v$ is $1-e^{-\tau_S(h)}$ where $\tau_S$ is the Sobolev Optical depth \citep{Lamers1999}.  Then, accounting for absorption by the outflow, we have 
\begin{eqnarray}
&\resizebox{0.41\textwidth}{!}{$
I(h)_{\rm{abs,blue}}/I_0 = \frac{F_{\lambda}}{F_{c,\lambda}} -  \frac{F_{\lambda}}{F_{c,\lambda}} \frac{\int_0^{1}2\pi h(1-e^{-\tau_S(h)})dh}{\pi (1)^2}$}.
\label{absorption_component1}
\end{eqnarray}

\noindent To stay consistent with the old SALT model, we want to write the new integral in terms of the parameters $x = v_{obs}/v_0$ and $y = v/v_0$.  In this context, $h = [y^{2/\gamma}-x^2y^{2(1-\gamma)/\gamma}]^{1/2}$ (i.e., $h$ equals the $\xi$ coordinate of $\Gamma_x$, see Appendix 2 of \citealt{Carr2018} for a calculation).  Finally, the value of the normalized absorption spectrum becomes
\begin{equation}
\begin{aligned}
&I(x)_{\rm{abs,blue}}/I_0 =&\\[1em]
&\resizebox{0.41\textwidth}{!}{$\frac{F(x)}{F_c(x)} -\frac{2}{\gamma}\frac{F(x)}{F_c(x)} \int_{\rm{max}(x,1)}^{y_1}(y^{(2-\gamma)/\gamma}-(1-\gamma)x^2y^{(2-3\gamma)/\gamma})(1-e^{-\tau_S(x,y)})dy$,}&\\
\end{aligned}
\label{absorption_component2}
\end{equation}
where an explicit formula for $\tau_S$ is provided below.  The upper bound of integration, $y_1$, represents the highest intrinsic velocity a shell can have and still contribute to continuum absorption (see \citealt{Carr2018} for an explicit formulae to calculate $y_1$).

\begin{figure}
  \centering
\includegraphics[scale=0.5]{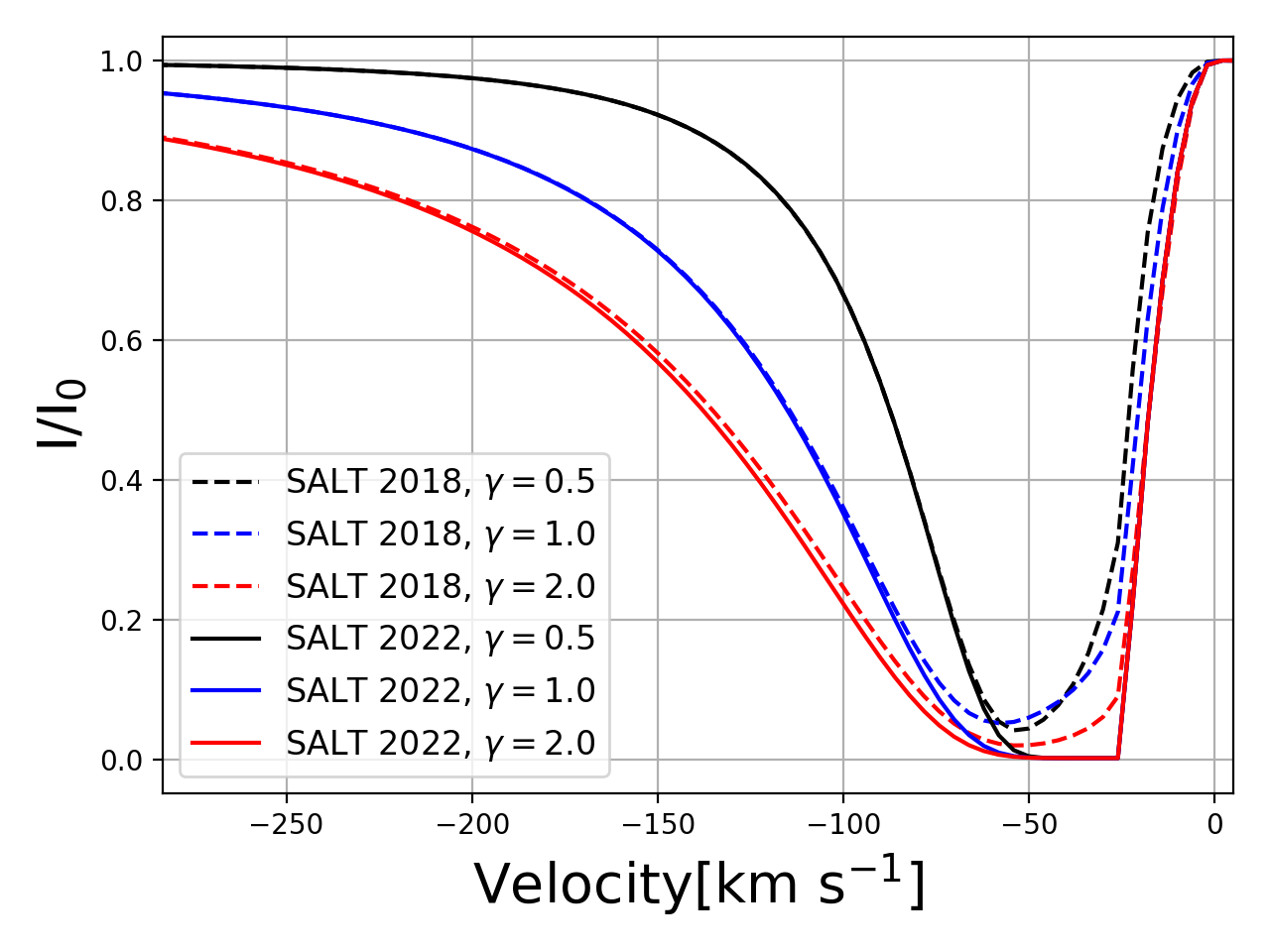}
 \caption{Absorption profile comparison between the 2018 (dashed) and 2022 (solid) versions of SALT for $\gamma = 2$ (blue), $\gamma = 1$ (red), and $\gamma=.5$ (violet).  The largest differences between the two models occur near zero observed velocity because the two models converge at large radii.  The rest of the parameters used in these models are $\tau_0 = 330$, $\gamma = 1.0$, $v_0 = 25 \ \rm{km \ s^{-1}}$, $v_{\infty} = 450 \ \rm{km \ s^{-1}}$, $v_{ap} = 450 \ \rm{km \ s^{-1}}$, $f_c = 1$, and $\kappa = 0$.}
   \label{f3}
\end{figure}

We have plotted absorption line profiles for various values of $\gamma$ using both the new and old SALT model in Figure~\ref{f3}.  The new models (solid lines) reach saturation closer to the continuum source compared to the older models (dashed lines). This is reflected in the departures of the old/new profiles at low velocities. Because the different models largely resemble one another at larger radii, the profiles at large velocities agree with each other. The latter reflects the fact that both the surface of constant observed velocity, $\Omega_v$, and the portion of the shell covering the source, appear flat at large radii.  Under such conditions the two different versions of the SALT model are identical. 

\subsection{New Emission Profile}

\begin{figure*}
  \centering
\includegraphics[scale=.8]{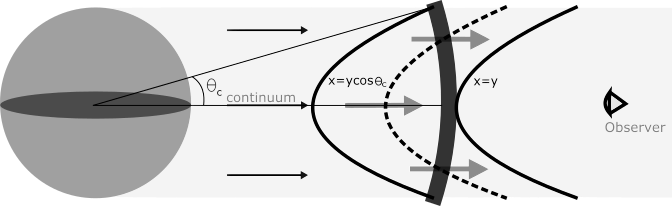}
 \caption{Same base setup as Figure~\ref{f2}.  The structure on the far right represents a shell centered on the source.  Three different surfaces, $\Omega_{x_i}(y)$, of constant observed velocity, $x_i$, in the range $[y \cos{\Theta_C},y]$ are shown.  Notice that each surface intersects the shell along the continuum, and will therefore absorb continuum photons at different wavelengths (i.e., the wavelength at resonance with material moving at observed velocity, $x_i$).  To compute the amount of energy absorbed by the shell, we want to add up all the energy absorbed from each surface by the shell as we continuously deform from one surface to the next.}
 \label{f4}
\end{figure*} 

To stay consistent with our new derivation of the absorption profile, we must also account for the absorption of non-radially traveling photons by a shell before reemission\footnote{The line profile resulting from the resonant scattering of photons through a spherically symmetric outflow is known to have equal absorption and emission equivalent widths (i.e., assuming energy is conserved, \citealt{Prochaska2011}).  Thus the greater absorption achieved by the new absorption component of SALT (see Figure~\ref{f3}) suggests a similar adjustment must be made to the emission component as well.}.  As a first step, we seek to compute the amount of energy absorbed by a single shell of intrinsic velocity, $y$, and thickness, $dy$, from the continuum (i.e., we consider only photons emitted parallel to the line of sight).  In other words, we aim to compute the amount of energy absorbed from the continuum by the surfaces of constant observed velocity, $\Omega_x(y)$, which intersect the shell in the observed velocity range, [$y\cos{\Theta_C},y$], where $\Theta_C = \arcsin{y^{-1/\gamma}}$ is the angle subtended by the shell along the continuum.  A diagram has been provided in Figure~\ref{f4}.  We can view the problem geometrically by treating $\Omega_{x}(y)$ as a homotopy, or as a continuous map from one surface to the next, where our goal is to add up the varying fractions of continuum energy removed by each surface intersecting the shell as we continuously deform from $x = y\cos{\Theta_C}$ to $x = y$.  Once again, it will be sufficient to consider the curve or intersection of each surface with the $s\xi$-plane, $\Gamma_x(y)$.  Fortunately, we have already computed the fraction of the continuum energy absorbed by $\Gamma_x(y)$ in the interval $dy$ in the integrand of Equation~\ref{absorption_component2}.  Thus, the amount of continuum energy absorbed by material in the shell at resonance with observed velocity $x \in [y\cos{\Theta_C},y$] is 
\begin{equation}
\begin{aligned}
&dL_{x,\rm{c,shell}} = &\\[1em]
&\resizebox{0.41\textwidth}{!}{$\frac{2}{\gamma}L_C(y^{(2-\gamma)/\gamma}+(\gamma-1)x^2y^{(2-3\gamma)/\gamma})(1-e^{-\tau_S(x,y)})dy$,}&\\
\end{aligned}
\label{shell_em}
\end{equation}  
where $L_C(x)$ is the amount of continuum energy emitted at resonance with material moving at velocity, $x$.  It follows that the total amount of continuum energy absorbed by the shell is 
\begin{equation}
\begin{aligned}
&L_{\rm{c,shell}} = \int \frac{dL_{x,\rm{c,shell}}}{dy}dx = \\[1em]
& \resizebox{0.41\textwidth}{!}{$\int^{y}_{y\cos{\Theta_C}}\frac{2L_c(x)}{\gamma}(y^{(2-\gamma)/\gamma}+(\gamma-1)x^2y^{(2-3\gamma)/\gamma})(1-e^{-\tau_S(x,y)})dx$.}&\\
\end{aligned}
\label{shell_em_c}
\end{equation}

Our goal is to find the total amount of energy absorbed by the shell, but Equation~\ref{shell_em_c} only considers rays emitted parallel to the line of sight.  Observe, however, that Equation~\ref{shell_em} has the form $dL_{x,\rm{c,shell}} = L_c*[\rm{fraction}]$ (see Equations~\ref{absorption_component1} and \ref{absorption_component2}) - that is, Equation~\ref{shell_em} computes the fraction of continuum energy, at a given wavelength, absorbed by the shell.  Furthermore, notice that any ray emitted by the source can be observed as continuum emission following a suitable rotation of reference frame.  (Just draw a line parallel to the emitted ray through the origin, then use this line to define the new line of sight.)  It follows that we can replace $L_c$ in Equation~\ref{absorption_component2} by $L(x)$, or the total energy emitted by photons at the resonant wavelength, because the fraction of the total energy absorbed by the shell is the same as that defined in Equation~\ref{shell_em}.  Therefore, the total energy absorbed by the shell becomes 

\begin{equation}
\begin{aligned}
&L_{\rm{shell}} = \\[1em]
& \resizebox{0.41\textwidth}{!}{$\int^{y}_{y\cos{\Theta_C}}\frac{2L(x)}{\gamma}(y^{(2-\gamma)/\gamma}+(\gamma-1)x^2y^{(2-3\gamma)/\gamma})(1-e^{-\tau_S(x,y)})dx$.}&\\
\end{aligned}
\end{equation}

The last step is to compute the line profile in terms of the observed velocities assuming isotropic reemission from the shell.  To accomplish this, we use the same emission band contour theory first developed by \cite{Beals1931} and used by \cite{Carr2018}.  We consider a band on the shell of area $2\pi r \sin{\theta}rd\theta$ where $r$ is the radius of the shell and $\theta$ is the angle subtended by the radius of the band (see \citealt{Carr2018}, Figure 4).  Since the source emits isotropically, we can assume that the energy absorbed by the band, $dL_{BC}$, is uniform - that is,

\begin{eqnarray}
dL_{BC}&=& L_{\rm{shell}}\frac{2\pi r^2 \sin{\theta} d\theta}{4\pi r^2}.
\end{eqnarray}   
Using the relation $dv_{\rm{obs}} = v \sin{\theta}d\theta$, we can rewrite the expression in terms of the observed velocities as
\begin{eqnarray}
dL_{BC}&=& L_{\rm{shell}}\frac{2\pi r^2 \sin{\theta} dv_{\rm{obs}} }{4\pi r^2 v} \\
&=& L_{\rm{shell}}\frac{dv_{\rm{obs}} }{2v}.
\end{eqnarray} 
 Assuming the band reemits isotropically, the observed flux emitted by the band, $dF_{BC}$, becomes 
 \begin{eqnarray}
 dF_{BC} = L_{\rm{shell}}\frac{dv_{\rm{obs}} }{4\pi r_{\infty}^2 2v},
 \end{eqnarray}
where $r_{\infty}$ is the distance to the observer.  Finally, after integrating over all shells, projecting, and normalizing by the continuum, the normalized blue emission profile becomes   

\begin{equation}
\begin{aligned}
&I(x)_{\rm{em,blue}}/I_0= \int \frac{1}{F}\frac{dF_{BC}}{dv_{\rm{obs}}}dv = &\\[1em]
&\resizebox{0.41\textwidth}{!}{$\int_{\rm{max}(x,1)}^{y_{\infty}}\frac{dy}{2y}\int^{y}_{y\cos{\Theta_C}}\frac{2}{\gamma}\frac{F(x^{\prime})}{F_c(x)}(y^{(2-\gamma)/\gamma}+(\gamma-1){x^{\prime}}^2y^{(2-3\gamma)/\gamma})(1-e^{-\tau_S(x^{\prime},y)})dx^{\prime}$,}&\\
\end{aligned}
\label{em1}
\end{equation}
and the normalized red emission profile becomes 
\begin{equation}
\begin{aligned}
&I(x)_{\rm{em,red}}/I_0= &\\[1em]
&\resizebox{0.41\textwidth}{!}{$\int_{y_1}^{y_{\infty}}\frac{dy}{2y}\int^{y}_{y\cos{\Theta_C}}\frac{2}{\gamma}\frac{F(x^{\prime})}{F_c(x)}(y^{(2-\gamma)/\gamma}+(\gamma-1){x^{\prime}}^2y^{(2-3\gamma)/\gamma})(1-e^{-\tau_S(x^{\prime},y)})dx^{\prime}$,}&\\
\end{aligned}
\label{em2}
\end{equation}
where $F(i) = L(i)/4\pi r_{\infty}^2$ represents the flux at observed velocity, $i$.  The red and blue components account for reemission occurring at negative and positive observed velocities, respectively.  The only difference between the two profiles is the range of integration.  The lower bound for the red integral, $y_1$, takes the same definition as in the absorption case and accounts for the blocking of rays emitted from behind the source.  Lastly, the techniques developed by \cite{Carr2018,Carr2021} to account for a biconical outflow geometry, a limiting observing aperture, a dusty CGM, and holes in the outflow still apply.  Furthermore, the multiple scattering procedure of \cite{Scarlata2015} to account for fluorescent and resonant scattering also carries over to this model in the obvious way.   

\begin{figure}
  \centering
\includegraphics[scale=0.5]{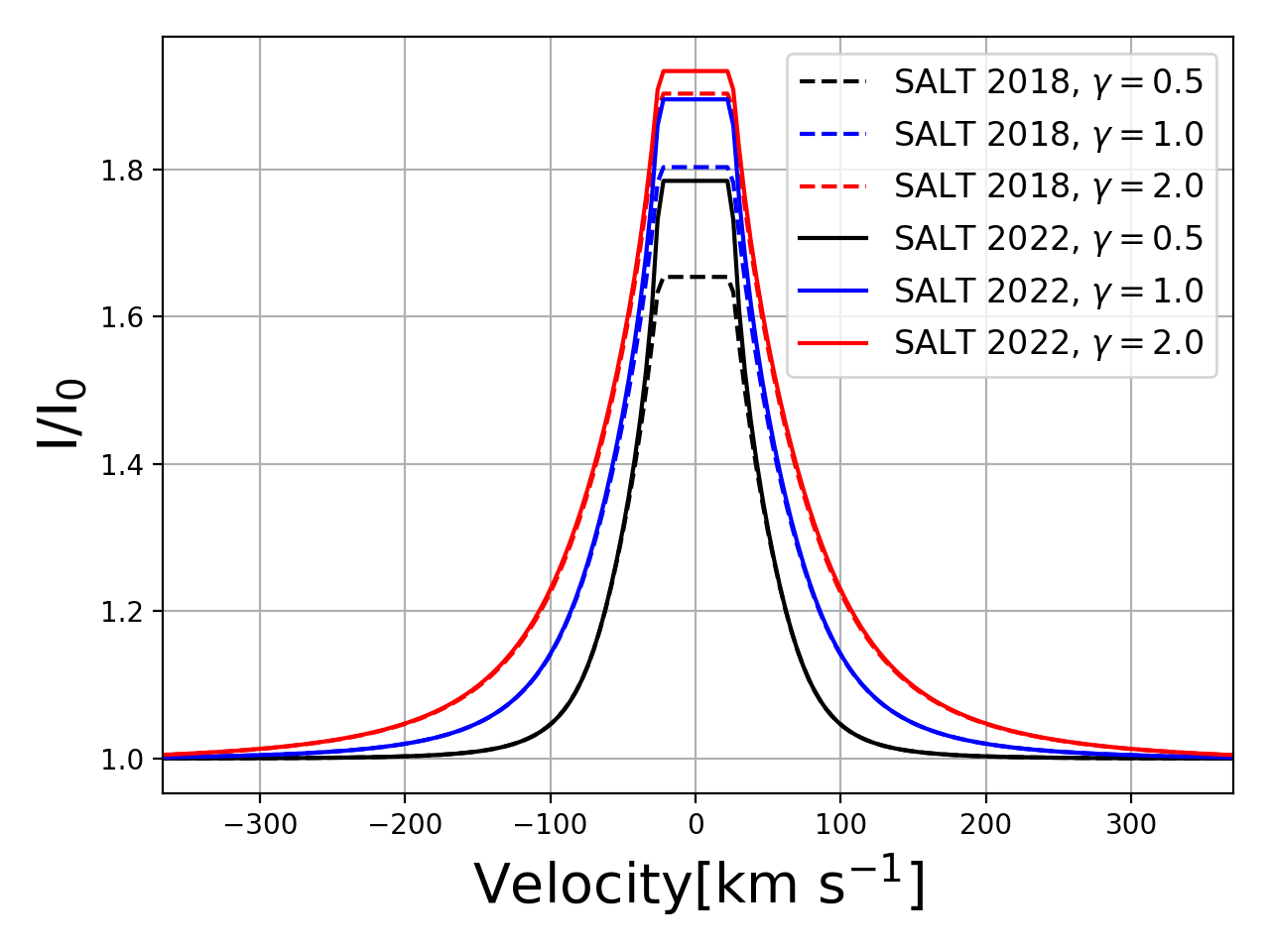}
 \caption{Same as Figure~\ref{f3}, but for the emission case.  Similar to the absorption case, the differences between the two models are largest at low observed velocities.  This is because the radiation transfer problem approaches that of a point source at large radii.  The rest of the parameters used in these models are $\log{(N \ [\rm{cm}^{-2}])} = 15$ (assumes $R_{SF} = 1 \ \rm{kpc}$), $\gamma = 1.0$, $v_0 = 25 \ \rm{km \ s^{-1}}$, $v_{\infty} = 450 \ \rm{km \ s^{-1}}$, $v_{ap} = 450 \ \rm{km \ s^{-1}}$, $f_c = 1$, and $\kappa = 0$.   }
   \label{f5}
\end{figure}

Using an identical setup as in Figure~\ref{f3}, we have plotted emission profiles for various values of $\gamma$ using both the new and old SALT model in Figure~\ref{f5}.  Similar to the absorption case, the new model achieves more emission at smaller observed velocities, and converges to the old model at higher observed velocities.  This time, however, the latter can be explained by noting that the behavior of the line profile should approach that of a point source as the radius of the shell grows large.  

\subsection{Arbitrary Density Field}

Here we relax the assumption of a constant mass outflow rate assumed by, for example, \cite{Carr2018,Carr2021}, and assume a general density field to characterize the flow.  We assume the density field, $n(r)$, follows a power law of the form $n(r) = n_0(r/R_{SF})^{-\delta}$ (see \citealt{Werk2013}), where $n_0$ is the number density of the relevant ion at $R_{SF}$.  The density field determines the strength of the absorption coefficient and is manifestly part of the optical depth.  Following \cite{Carr2018}, we compute the Sobolev optical depth \citep{Castor1970,Lamers1999} as     
\begin{eqnarray}
\tau_S(r)&=& \frac{\pi e^{2}}{mc}f_{lu}\lambda_{lu}\left[1-\frac{n_ug_l}{n_lg_u}\right]n(r)\frac{r/v}{1 + \sigma  \cos^2{\phi}},\label{optical_depth}
\label{tau}
\end{eqnarray}
where $\sigma = \frac{d \ln(v)}{d \ln(r)} -1$ and $\phi$ is the angle between the velocity and the trajectory of the photon.  $f_{lu}$ and $\lambda_{lu}$ are the oscillator strength and wavelength for the $lu$ transition, respectively.  All other quantities take their usual definition.  Neglecting stimulated emission (i.e., $\left[1-\frac{n_ug_l}{n_lg_u}\right] = 1$) and writing Equation~\ref{tau} in terms of the quantities relevant to the SALT model, the expression for the optical depth in Equations~\ref{absorption_component2}, \ref{em1}, and \ref{em2} becomes
\begin{eqnarray}
\tau_S(x,y) &=& \frac{\tau_0}{1+(\gamma-1)\left(x/y\right)^2}y^{(1-\gamma-\delta)/\gamma},
\label{tau_s}
\end{eqnarray}
where
\begin{eqnarray}
\tau_0 = \frac{\pi e^2}{mc} f_{lu}\lambda_{lu} n_0 \frac{R_{SF}}{v_0}.
\end{eqnarray}

\begin{figure}
  \centering
\includegraphics[scale=0.34]{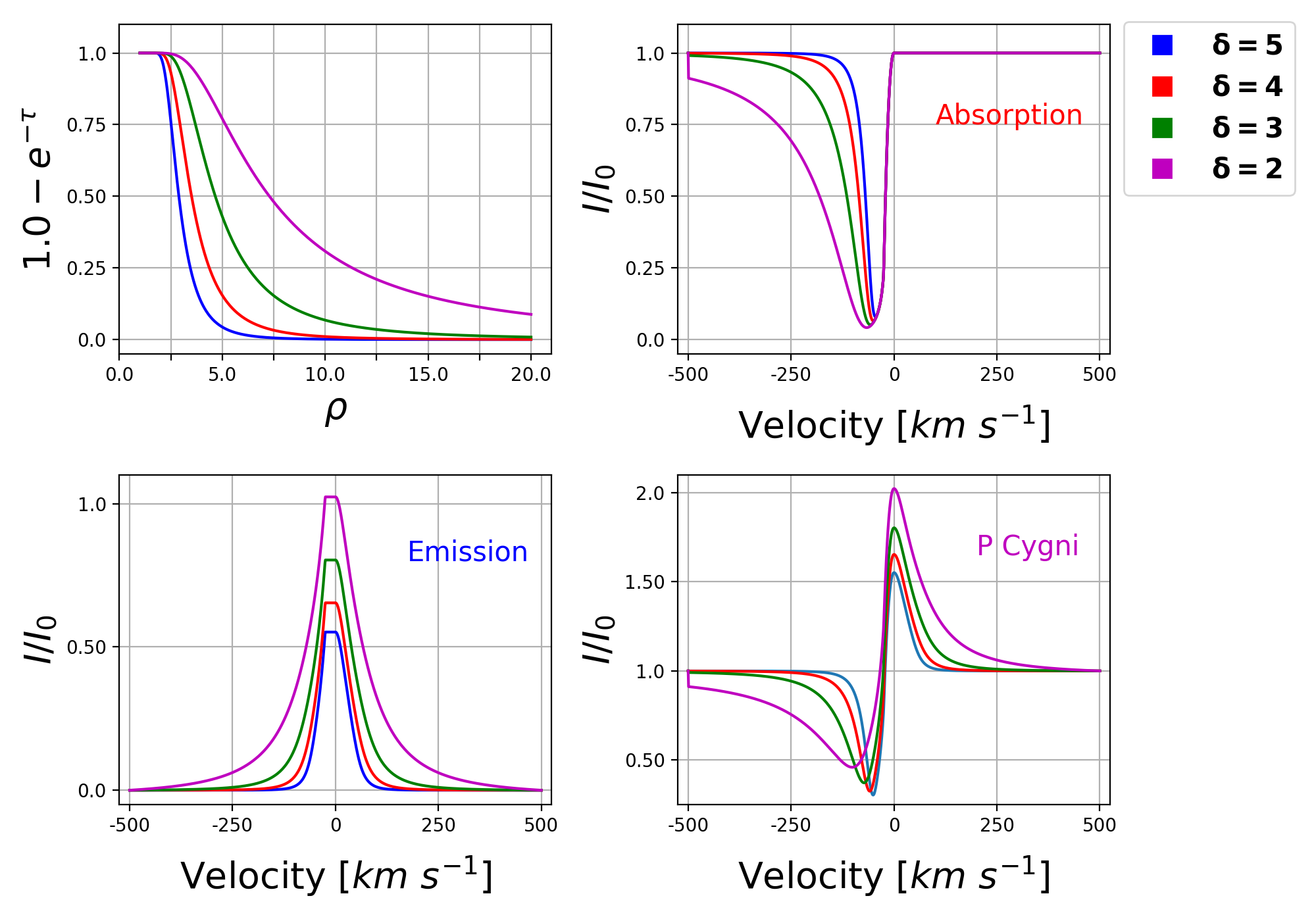}
 \caption{Starting from the top left.  The fraction of energy, $1-e^{-\tau_S(\rho)}$, absorbed by a shell of radius, $\rho$, plotted against the entire extent of the outflow ($1<\rho<20$).  As $\delta$ increases, a larger volume of energy is removed from the continuum.  To the right are the absorption profiles and bottom left the emission profiles.  Each feature increases with decreasing $\delta$ as expected.  The bottom right shows the complete P Cygni profiles.  The rest of the parameters used in this model are $\log{(N \ [\rm{cm}^{-2}])} = 15$, $\gamma = 1.0$, $v_0 = 25 \ \rm{km \ s^{-1}}$, $v_{\infty} = 500 \ \rm{km \ s^{-1}}$, $v_{ap} = 500 \ \rm{km \ s^{-1}}$, $f_c = 1$, and $\kappa = 0$.  }
   \label{f6}
\end{figure}

We investigate how the line profile varies with $\delta$ for a spherical outflow with fixed column density, $\log{(N \  [\rm{cm}^{-2}])} = 15$, in Figure~\ref{f6}.  Identical colors correspond to the same model in each panel.  The fraction of the continuum, $1-e^{-\tau(\rho)}$, absorbed by a shell at radius, $\rho$, is shown in the upper left panel.  The absorption, emission, and full P Cygni profile are shown in the upper right, bottom left, and bottom right panels, respectively.  In this model, a constant mass outflow rate is achieved for $\delta = 3$ - that is, for $\delta = \gamma + 2$ (see \citealt{Carr2018}) where $\gamma = 1$ is the power index of the velocity field.  For $\delta < 3$, the density field decays slower than that required for a constant mass outflow rate and leads to more absorption and more emission.  Moving in the opposite direction, for $\delta > 3$, the density field decays faster than that required for a constant mass outflow rate and leads to less absorption and less emission. 

The rest of this paper is dedicated to testing the upgraded version of SALT.  As such, all of the following results which depend on SALT can be assumed to have been derived from the new model.  For convenience, we have provided a complete list of the SALT parameters along with their definitions in Table~\ref{tab:pars}.

\section{RASCAS Upgrade: Adaptive Mesh Refinement}
\label{AMR}

RASCAS \citep[for RAdiation SCattering in Astrophysical Simulations,][]{Michel-Dansac2020} is a numerical radiation transfer code to compute the propagation of (non)resonant photons through astrophysical environments. These may be provided by numerical simulations, e.g. of galaxies \citep{Mauerhofer2021,Garel2021} or turbulent molecular clouds \citep{Kimm2022}, or they may be an implementation of idealised models. RASCAS treats the radiation transfer process in full, but - as a numerical routine - must rely on a 3-Dimensional grid to perform computations.  

While a regular grid may be enough to describe a number of idealised models with enough accuracy \citep[e.g.][]{Verhamme2006, Prochaska2011, Song2022}, this approach quickly demands prohibitive amounts of memory in the general case. The models we wish to explore, in particular, feature steep velocity and density gradients which require very high resolution. In the present paper, we use an updated version of the RASCAS code, which allows us to implement idealised models with adaptive mesh refinement (AMR). This adaptive grid is defined as follows.

The user first fixes a maximum level of refinement, $l_{\rm max}$, which defines the maximum grid size, corresponding to $2^{l_{\rm max}}$ cells on a side.  

The refinement procedure then begins as a loop on cells existing at each refinement level, starting with the unique cell at level $l=0$ (the only one always defined at initialisation) to possible cells at level $l=l_{\rm max}-1$. At each level, we decide to refine a cell if its properties meet any of the following criteria:
\begin{itemize}
    \item density variations across the cell are such that $Q_\rho = (\rho_{\rm max} - \rho_{\rm min}) / (\rho_{\rm max} + \rho_{\rm min}) > \epsilon_\rho$, where $\rho_{\rm max}$ and $\rho_{\rm min}$ are the maximum and minimum densities found in the cell, and $\epsilon_\rho \leq 1$ is a user-defined parameter. In order to speed up computations, we first evaluate $\rho_{\rm max}$ and $\rho_{\rm min}$ at grid points of level $l+1$, and only if $Q_\rho < \epsilon_\rho$ do we evaluate points at level $l+2$, and so on until the cell is refined or no variation is found at a resolution of $l_{\rm max}$. 
    \item velocity variations across the cell are such that $Q_v > \epsilon_v$, where $Q_v$ is defined as $Q_\rho$ but using the norm of the velocity field instead of density.
    \item the maximum velocity difference across the cell ($\Delta v = v_{\rm max}-v_{\rm min}$) is larger than the thermal (or turbulent) velocity dispersion $v_{\rm th}$. $\Delta v$ is evaluated along $Q_v$ with the same strategy.
\end{itemize}
Note that our strategy will robustly identify (and refine if needed) details down to a scale $1/2^{l_{\rm max}}$. 

If one of the criteria is met, the cell is refined and 8 level $l+1$ cells are inserted in a linked list to be later tested. 

After this first pass, we run a second pass over all cells at all levels in order to ensure that two adjacent cells never differ by more than one level of refinement. This rule is inherited from the AMR code RAMSES \citep{Teyssier2002}, and the search for neighboring cells in RASCAS requires it \citep[see][for more details]{Michel-Dansac2020}. 

\begin{figure}
  \centering
\includegraphics[scale=0.42]{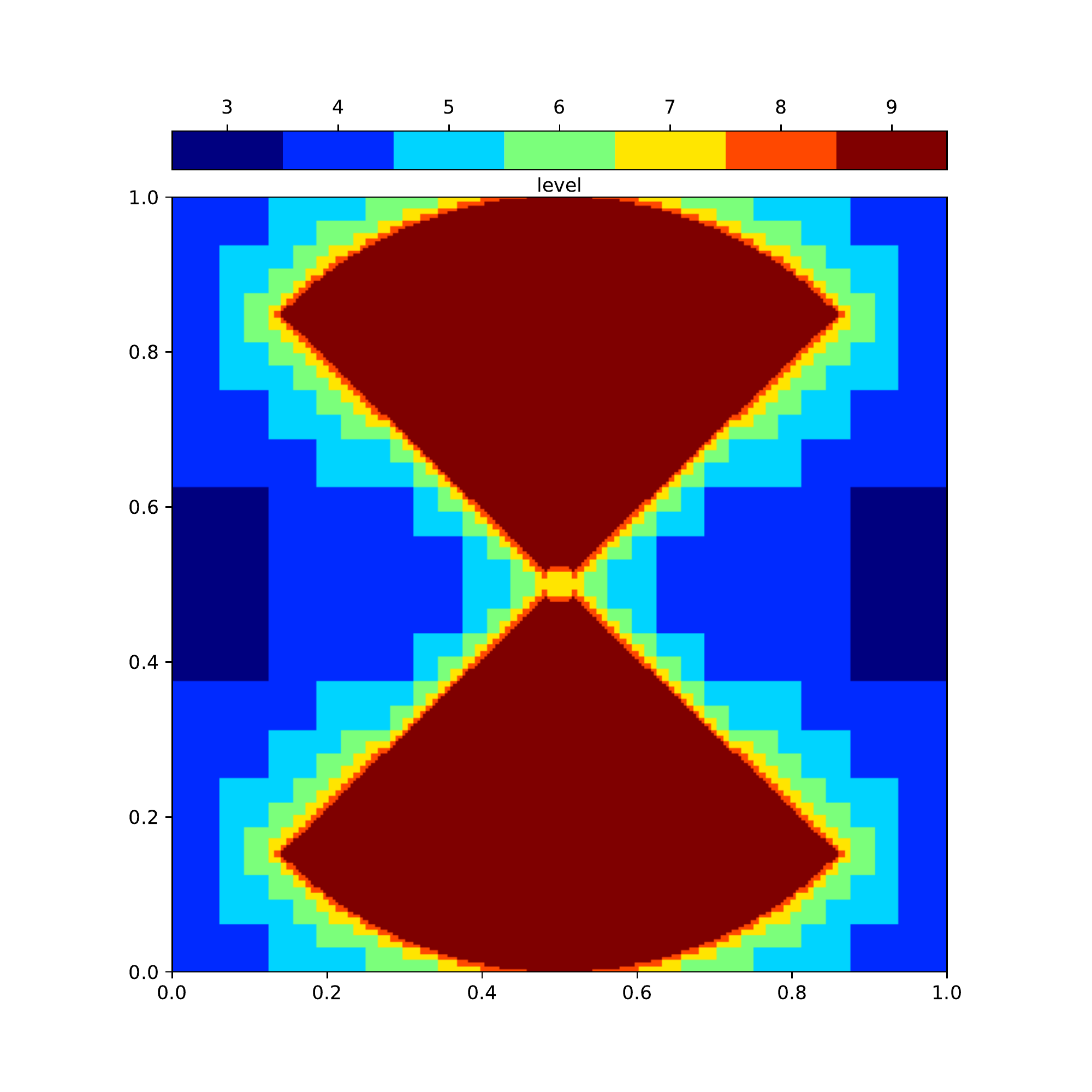}
 \caption{Refinement map of the biconical wind model. Colors indicate the maximum level of refinement reached along the line of sight through a thin, centered slice of the full 3 dimensional grid generated with RASCAS with the new AMR scheme.}
   \label{f7}
\end{figure}

Generating an AMR grid tailored to each idealised model is an improvement, but in practice, this is not enough to keep the memory footprint of the code low for any model.
An example of an AMR grid for the biconical wind model is shown in Figure~\ref{f7}. As it can be seen, the velocity/density gradients and the small value of thermal velocity ($v_{\rm th} = 1\ \rm{km} \ \rm{s}^{-1}$) used for this model makes the whole wind region resolved at the maximum level of refinement. 

We thus also use domain decomposition, as described in \citet{Michel-Dansac2020}, and build the AMR grid across any number of pre-defined sub-domains. Because some of our models break spherical symmetry, we have introduced into RASCAS a new implementation of multi-domain peeling off \citep[see][]{Dijkstra2017}, which allows us to compute efficiently mock spectra in any direction despite this domain decomposition. 

Once the mesh is generated, the propagation of the photon packets is computed in the standard way with RASCAS, as described in detail in \citet{Michel-Dansac2020}, and we have introduced no further modifications to the code. 

It is interesting to understand the resolution required with RASCAS to reach numerical convergence.  Suppose a photon emitted at frequency, $\nu$, comes into resonance at a point, $p$, moving with the outflow at a velocity, $v$, with respect to the frame in which the photon was emitted. If we ignore line broadening, the optical depth will be effectively zero everywhere except exactly at $p$ - that is, the photon will only come into resonance and interact with the outflow exactly at $p$.  In this circumstance, the point $p$ is often called the Sobolev point (see \citealt{Lamers1999}) and the profile function or cross section will be a Dirac delta function.  If we consider line broadening, the photon will be able to come into resonance in a volume centered on $p$ where the thermal motion can negate the Doppler shift due to the bulk flow of the outflow.  This volume describes where a resonance interaction between the outflow and photon can occur and is called the line interaction region (see \citealt{Lamers1999}).  The line interaction region can be approximated by the Sobolev length, $S_L$, or the distance over which the velocity field changes by an amount, $v_{th}$. The volume enclosed by computing $1.5 S_L$ around $p$ has been found to be a good approximation to the line interaction region \citep{Lamers1999}.  Therefore, to properly resolve resonance scattering in a moving medium, RASCAS must at least resolve the interaction region. 

\begin{figure}
  \centering
\includegraphics[scale=0.53]{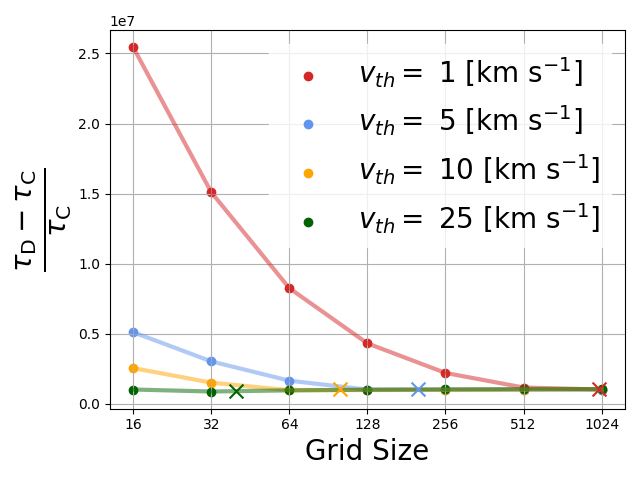}
 \caption{Normalized difference between the numerically derived optical depth, $\tau_{D}$, and the exact analytical expression or the continuum limit, $\tau_{C}$, versus grid size.  The different colors represent the different values of the thermal velocity, $v_{th}$.  The hash marks or 'Xs' represent the locations where the Sobolev length equals the size of a single grid cell - that is, where the interaction region can be resolved by the grid.  }
 \label{f8}
\end{figure}

In Figure~\ref{f8}, we test for convergence of the numerically derived optical depth, $\tau_D$, with the true (continuous) optical depth, $\tau_C$, inferred from the underlying idealized model as a function of grid size (i.e., the dimension of the grid).  We have marked each location where the grid size should be large enough to resolve the interaction region - that is, where

\begin{eqnarray}
S_L > \frac{2R_W}{\rm{Grid \ Size}}.
\end{eqnarray}
It is clear from the figure that the optical depth has successfully converged at these locations as expected.

\section{Testing SALT With RASCAS}
\label{lowthermalvelocitytests}

\begin{figure*}
  \centering
\includegraphics[scale=0.59]{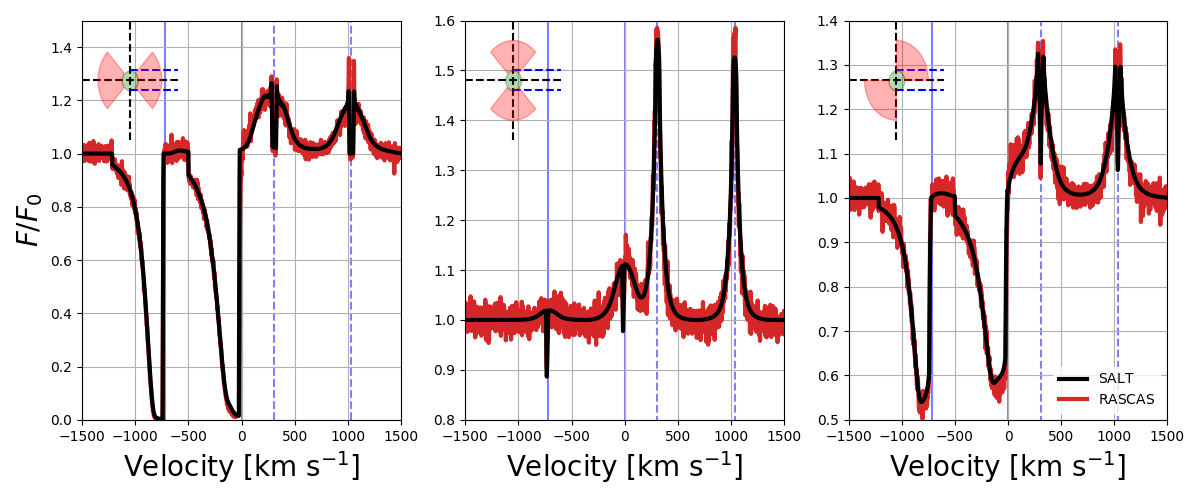}
 \caption{SALT (red) and RASCAS (black) predictions for the Si II 1190\AA,1193\AA\ doublet line profiles for a biconical outflow ($\alpha = 45^{\circ}$).  The different panels represent different orientations of the outflow with respect to the observer ($\psi = 0^{\circ}$,$90^{\circ}$,$45^{\circ}$ for the \textbf{\emph{Left}}, \textbf{\emph{Middle}}, \textbf{\emph{Right}} \textbf{\emph{Panels}}, respectively); all emblems in the upper left corner are to be viewed from the right.}
 \label{f9}
\end{figure*}
       
The SALT formalism (see Section~\ref{Formalism}) relies on approximations (e.g., the Sobolev approximation) and idealized model set ups to solve the radiation transfer equation.  In this context, SALT is capable of exploring large dimensional parameter spaces quickly at minimal cost in terms of memory and storage space, making it an efficient means for interpreting data (e.g., \citealt{Carr2021}).  In contrast, RASCAS relies on Monte Carlo numerical calculations to solve the radiation transfer equation in full.  While accurate and highly adaptable, this approach is costly and renders RASCAS an inefficient means for constraining models in large parameter spaces.  When used together, however, these models have the potential to create a powerful tool to perform precision astronomy (e.g., Carr et al. 2023, in prep), but they must be shown to be both compatible and well tested.  While RASCAS has been tested and upheld to various known analytical results throughout the literature (see \citealt{Michel-Dansac2020}), the SALT model has yet to be thoroughly tested.  In this section, we seek to thoroughly test SALT by comparing model predictions to those of RASCAS directly and then assuming the agreement with other models in the literature follows transitively.  

To form a rigid basis of comparison, we attempt to recreate the exact same radiation transfer experiment assumed by SALT in RASCAS - that is, using the idealized models module (see \citealt{Michel-Dansac2020}), we construct a biconical outflow defined by the same density and velocity fields as SALT in RASCAS.  Photons are drawn uniformly and emitted isotropically from a centrally located spherical source of radiation, and isotropic scattering is further assumed to occur in the outflow.  See Figure~\ref{f1}.  We set the Doppler parameter, $v_{th} = 1 \ \rm{km} \ \rm{s}^{-1}$, in RASCAS to ensure we are comparing SALT to RASCAS in the limit of low thermal velocity (i.e., in the limit that the Sovolev approximation is valid, see Section~\ref{sc}).  To test SALT against RASCAS, we first compare the predictions of SALT to RASCAS with a fiducial model, then isolate and change a single parameter between subsequent predictions.  The fiducial model assumes a spherical outflow (i.e., $\alpha = 90^{\circ}$), a homologous flow (i.e, $\gamma = 1.0$), a constant outflow rate (i.e., $\delta = \gamma + 2.0 = 3.0$), no dust (i.e., $\kappa = 0.0$), no holes or clumps (i.e., $f_c = 1.0$), a sufficiently large aperture to capture the entirety of the outflow (i.e., $v_{ap} > v_{\infty}$), a launch velocity, $v_0 = 25 \ \rm{km} \ \rm{s}^{-1}$, a terminal velocity, $v_{\infty} = 500 \ \rm{km} \ \rm{s}^{-1}$, and a column density, $\log(N_{Si^{+}} \ [\rm{cm}^{-2}]) = 15$.  For a complete list of the SALT parameters and their definitions, see Table~\ref{tab:pars}.  

For each comparison, we have chosen to model the Si II 1190\AA,1193\AA \ doublet absorption lines.  All relevant atomic information has been provided in Table~\ref{tab:atomicdata}.  These transitions have a fluorescent component (see \cite{Scarlata2015} for a description of the atomic structure) and will thoroughly test every aspect of the SALT model including the multiple scattering procedure developed by \cite{Scarlata2015}.  The following subsections are broken down according to the parameters tested.  All RASCAS predictions rely on the AMR routine described above to ensure all boundaries and gradients are resolved.  Specifically, we set the $l_{\rm max} = 10$, $\epsilon_\rho  = 1.0$, $\epsilon_v  = 1.0$, used the velocity refinement condition (i.e., $\Delta v > v_{th}$), and propagated 5 million photon packets. 

\subsection{Geometry}
SALT assumes a biconical outflow geometry characterized by a half opening angle, $\alpha$, and an orientation angle with respect to the line of sight, $\psi$.  To account for the effects of varying the opening and orientation angles on the predicted line profiles of SALT, \cite{Carr2018} first predicted the spectrum for a spherical outflow then removed energy from each shell according to a multiplicative factor which they called the geometric scale factor, $f_g$.  $f_g$ is a complicated function of $\alpha$ and $\psi$ which relies on the theory of band contours (see \citealt{Beals1931}), and as such, deserves to be tested.  In Figure~\ref{f9}, we compute the Si II 1190\AA,1193\AA\ doublet absorption profile for a biconical outflow with half opening angle, $\alpha = 45^{\circ}$.  Each panel shows the predicted spectrum of both SALT and RASCAS as it would appear from a different orientation with respect to the line of sight.  The agreement between RASCAS and SALT is impeccable for each orientation.  This not only reassures the accuracy of the calculations of \cite{Carr2018}, but also demonstrates the effectiveness of the AMR upgrade of RASCAS to resolve the boundary of the outflow.

\begin{figure*}
  \centering
\includegraphics[scale=0.59]{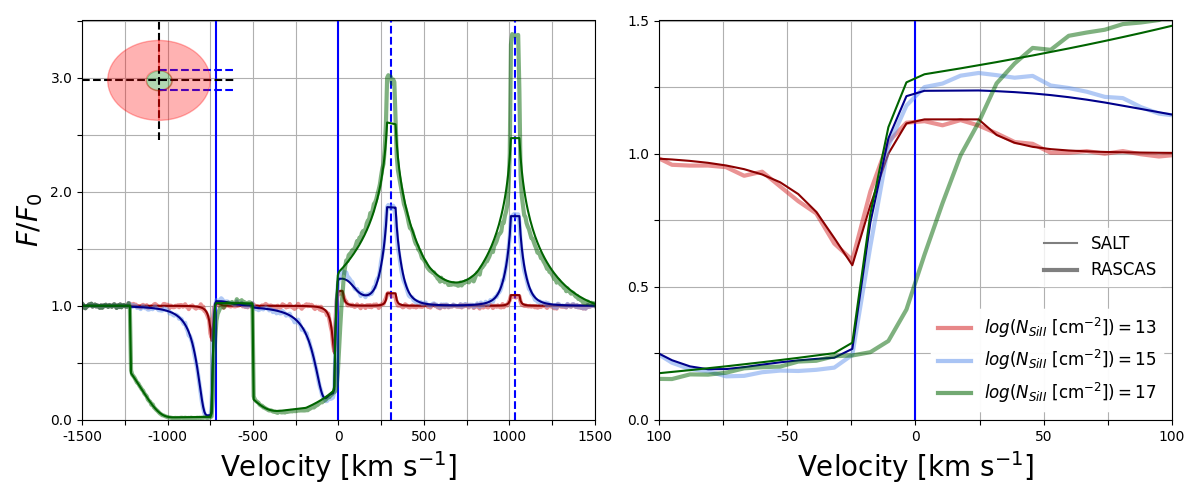}
 \caption{Spectral predictions of SALT and RASCAS for varying column densities.  The fiducial model is the same as Figure~\ref{f9}, except it now assumes a spherical outflow geometry.  \textbf{\emph{Left Panel}} SALT and RASCAS predictions for column densities $\log{(N_{Si^{+}} \ [\rm{cm}^{-2}])} = 15, 16, \ \rm{and} \ 17$ shown in red, blue, and green, respectively.  For $\log{(N_{Si^{+}} \ [\rm{cm}^{-2}])} = 17$, the predictions of RASCAS and SALT start to diverge. In particular, RASCAS achieves red shifted absorption and more fluorescent emission than SALT.       \textbf{\emph{Right Panel}}  Close up view of the $1193$\AA\ transition line.  It is clear that the equivalent width of the redshifted absorption feature in RASCAS grows with increasing density.}
 \label{f10}
\end{figure*}

\subsection{Density}
We compare the predictions of SALT and RASCAS for different densities/column densities of a spherical outflow in Figure~\ref{f10}.  In each model, we only change the column density, $N$, by changing the number density of Si$^{+}$, $n_{0,Si^{+}}$, at the launch radius while keeping the other parameters fixed.  (See the definition of $N$ in Section~\ref{Discussion}.)  We assume a constant mass outflow rate (i.e., $\delta = \gamma + 2.0$) in every model.  As the column density grows large ($\log{(N_{Si^{+}} \ [\rm{cm}^{-2}])} \sim 17$), the predictions of SALT and RASCAS start to diverge.  In particular, RASCAS is able to achieve absorption occurring to the red or right of line center (see right panel).  Furthermore, when this is the case, RASCAS achieves more emission at the fluorescent wavelengths (see left panel).  The later reflects the fact that the photons absorbed at resonance by these transitions always have a nonzero probability of being reemitted at at a longer or fluorescent wavelength (which is set by the Einstein coefficients of the relevant transitions, see \citealt{Scarlata2015} for a description).  Since there are less atoms in the excited ground state, fluorescent photons have a much greater chance of escaping the outflow.  Therefore, by increasing the density or number of interactions, one increases the probability that a given photon escapes the outflow at fluorescence.  In regards to the former, the absorption of photons to the red of line center is a clear indication of the finite width of the interaction cross section.  Indeed, red shifted absorption is impossible in SALT because it assumes the Sobolev approximation.  Even though we set the width of the interaction cross section to be small (we set $v_{th}=1 \ \rm{km} \ \rm{s}^{-1}$ in RASCAS for these tests), there is still a nonzero chance that a photon can be absorbed out of resonance with the bulk flow of the outflow due to the random or turbulent motion in the gas.  We will take a closer look at the role density plays in the spectral predictions of RASCAS and SALT in the next section where we examine precisely how RASCAS and SALT compute the optical depth.      

\begin{figure*}
  \centering
\includegraphics[scale=0.59]{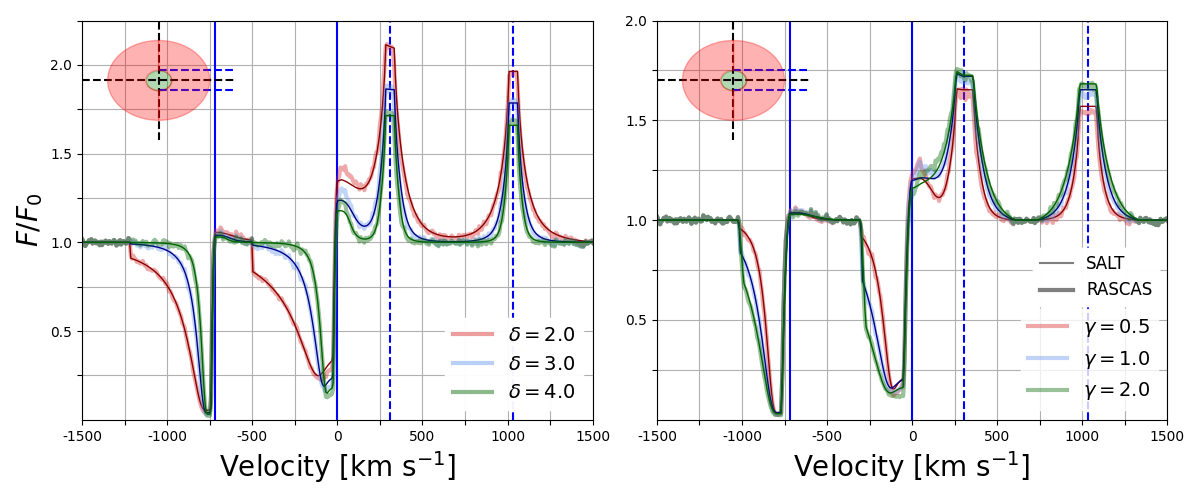}
 \caption{Spectral predictions of SALT and RASCAS for varying density and velocity field gradients.  The fiducial model is the same as Figure~\ref{f9}, except it now assumes a spherical outflow geometry.  \textbf{\emph{Left Panel}} Spectral predictions of SALT and RASCAS for different values of the power law index of the density field, $\delta$.  For these models, $\delta = 3.0$ represents a constant mass outflow rate.  \textbf{\emph{Right Panel}} Same as the left panel, except we vary the power law index of the velocity field, $\gamma$.  We assume a constant mass outflow rate for all models (i.e., $\delta = \gamma + 2.0$).  }
 \label{f11}
\end{figure*}

\begin{figure*}
  \centering
\includegraphics[scale=0.59]{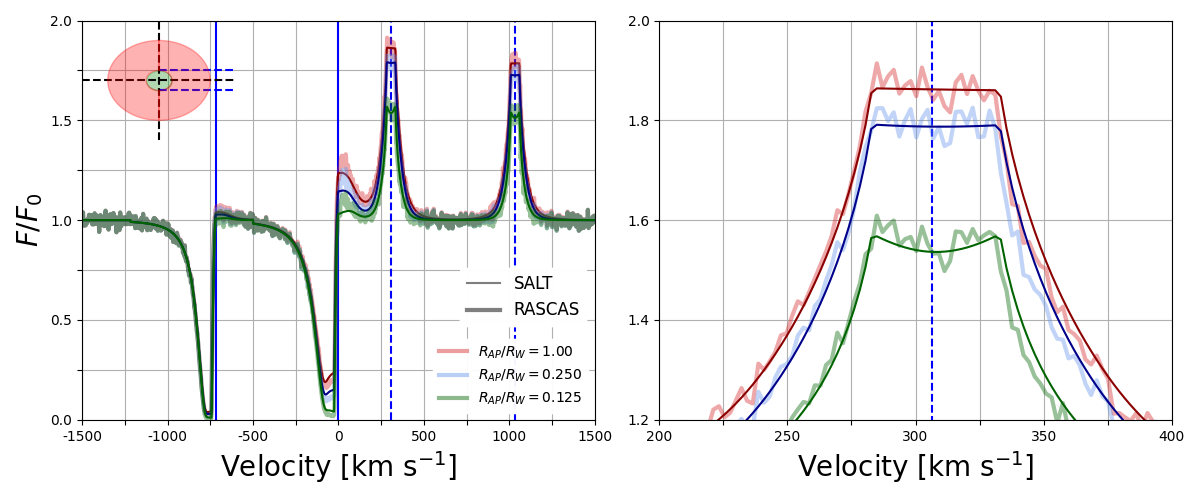}
 \caption{Spectral predictions of SALT and RASCAS for varying aperture sizes.  The fiducial model is the same as Figure~\ref{f9}, except it now assumes a spherical outflow geometry.  \textbf{\emph{Left Panel}} We test SALT against RASCAS for different sizes of the observing aperture ($R_{AP}$) in comparison to the wind ($R_W$).  \textbf{\emph{Right Panel}} Same as the left panel, except we have zoomed in on the $1194$\AA\ fluorescent peak.  The dip centered in the emission spike represents the loss of photons emitted by material flowing perpendicular to the line of sight (i.e., the material moving at zero observed velocity).}
 \label{f12}
\end{figure*}

\subsection{Density and Velocity Gradients}

We test SALT against RASCAS for different values of the density field gradient (left panel) and velocity field gradient (right panel) in Figure~\ref{f11}. For all models in the left panel, we assume a $\gamma = 1$ velocity field which implies $\delta = 3$ represents a constant mass outflow rate.  SALT is in excellent agreement with RASCAS for all values of $\delta$ shown.  For all models in the right panel, we assume a constant mass outflow rate for each trial (i.e., $\delta = \gamma + 2.0$, \citealt{Carr2018}).  Again, SALT and RASCAS are in excellent agreement for the range $0.5<\gamma<2.0$ which has been shown by \cite{Carr2021} and Huberty et al. (2023, in prep) to achieve reasonable fits to the spectra of local galaxies.  This agreement validates the generalization of SALT to arbitrary density fields made in section~\ref{Formalism}.  Furthermore, these results suggest that the Sobolev approximation and SALT are valid for the range of velocity field gradients covered as long as one remains reasonably close to a constant outflow rate.  We will explore this claim further in the next section.

\subsection{Observing Aperture}
Lastly, we test the SALT formalism for accounting for the effects of a limiting, finite circular observing aperture (e.g., the COS aperture) on a line profile.  In the formalism presented by \cite{Carr2018}, all reemission occurring in shells lying outside the aperture radius, as projected onto the plane of the sky, $R_{AP}$, is excluded from the line profile.  In this way, the amount of both blue-shifted and red-shifted emission is diminished as the aperture captures smaller and smaller portions of the outflow.  In contrast, the observing aperture is accounted for in RASCAS by simply adjusting the radius of the circular collecting area in the Monte Carlo scattering experiment.  In Figure~\ref{f12}, we compare the predictions of SALT and RASCAS for a spherical outflow with different aperture sizes.  The aperture scale is provided as a fraction of the wind radius, $R_W$.  We see that the predictions of both RASCAS and SALT are in near perfect agreement for every model.  The fact that this occurs on every scale tested demonstrates that the approximations of SALT agree with RASCAS on local scales, and does not represent a global averaging.  This fact bodes well for the application of SALT to integral field unit (IFU) data in future studies (e.g., those with JWST) where data can be collected from sub sets of the total image.

\begin{figure*}
  \centering
\includegraphics[scale=0.59]{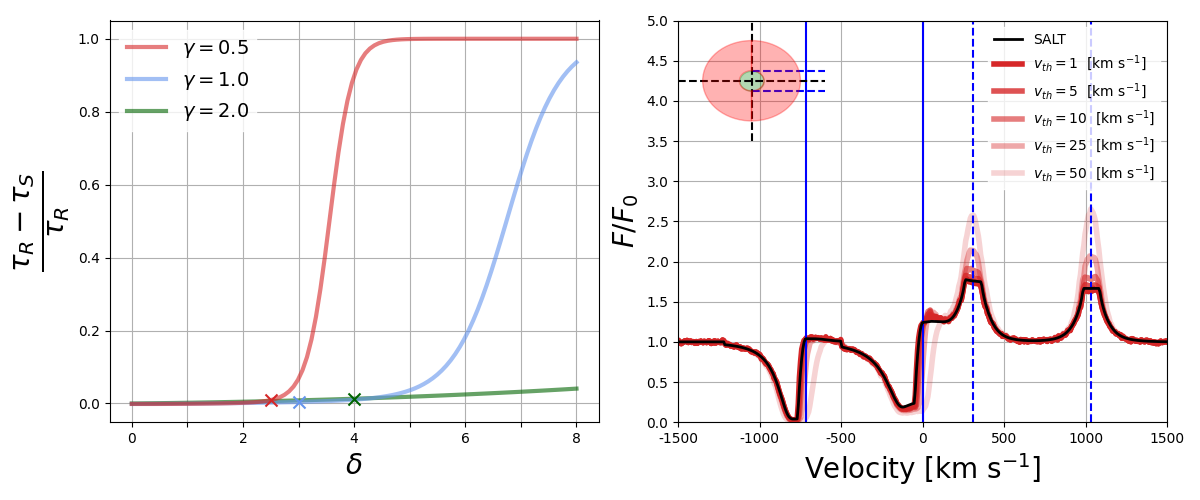}
 \caption{Testing the validity of the Sobolev criterion according to the Doppler parameter, $v_{th}$.  \textbf{\emph{Left Panel}}  Normalized difference between the true optical depth, $\tau_R$, as computed by RASCAS, with the Sobolev optical depth, $\tau_S$, as a function of the power law index of the density field, $\delta$.  Each curve assumes a different value for the power index of the velocity field, $\gamma = 0.5$ (red), $\gamma = 1.0$ (blue), and $\gamma = 2$ (green).  The optical depths were computed at a radius, $r = 200, 10, \sqrt{5} \ \rm{kpc}$ for each value of gamma, respectively, for an outflow with $R_{SF} = 1 \ \rm{kpc}$. The X's mark the locations for a constant outflow rate (i.e., $\delta = \gamma + 2.0$).  The remaining parameters required to compute the optical depth are $v_0 = 25 \ \rm{km} \ \rm{s}^{-1}$, $v_{\infty} = 500 \ \rm{km} \ \rm{s}^{-1}$, and $n_{0,Si^{+}} = 10^{-5} \ \rm{cm}^{-3}$.   \textbf{\emph{Right Panel}} Spectral predictions of RASCAS (red) and SALT (black) for different values of $v_{th}$ (SALT assumes $v_{th} = 0$).  The primary differences are that RASCAS can achieve redshifted absorption and more emission at fluorescence, all of which increase in equivalent width with increasing $v_{th}$.}
   \label{f13}
\end{figure*}

\section{Sobolev Criterion}\label{sc}

The SALT model assumes the Sobolev approximation \citep{Lamers1999} where radiation of a given wavelength can only interact with the outflow at a single point of resonance called the Sobolev point.  Consequently, SALT does not consider turbulence, thermal motion, or natural line broadening when computing the optical depth.  In contrast, RASCAS, which solves the radiation transfer process numerically, considers all three types of line broadening when computing the optical depth.  In this context, the optical depth can be computed as $\tau_R = \int n(r)\sigma \phi_{\nu} dr$ where $n(r)$ is the number density, $\sigma$ is the cross section for the relevant atomic interaction, and $\phi_{\nu}$ is the frequency dependent profile function.  The profile function is represented by the convolution of a natural Lorentzian profile and a Maxwellian velocity distribution for thermal broadening.  Its width is determined by the Doppler parameter, $v_{th}$ (see \citealt{Michel-Dansac2020} for an explicit definition).  Thus by adjusting the value of $v_{th}$ in different scattering experiments, we are able to explore how well the Sobolev approximation captures the radiation transfer process given different models of turbulence and bulk properties of the outflows.  Ultimately, since the purpose of SALT is to interpret real astronomical data, we want to be able to determine precisely when the Sobolev approximation is valid.  In this section, we review the theory behind the Sobolev approximation to establish a validity criterion within the SALT formalism and to better understand the differences between the spectral line predictions of RASCAS and SALT.     

The justification for the Sobolev approximation relies on the fact that in the case of a steep velocity field gradient, as one moves away from the Sobolev point, the surrounding material is quickly shifted out of resonance with the incoming photon and the absorption coefficient drops to effectively zero.  In this context, a scale length is typically assigned to the interaction region called the Sobolev length, $S_L$, which is defined as the distance from the Sobolev point over which the velocity field changes by an amount equal to the Doppler width - that is, $S_L = v_{th}/|d\mu v/dl|$, where $\mu = \cos{\theta}$ and $dl = \mu r$.  A typical requirement for the validity of the Sobolev approximation is for changes in the bulk flow to occur on length scales greater than the Sobolev length (e.g., \citealt{deKoter1993} and \citealt{Owocki1999}).  Using the density scale height, $H = n(r)/|d\mu n/dl|$, to define the flow scale, the criterion for the validity of the Sobolev approximation becomes $H >> S_L$, or in terms relevant to the SALT model,

\begin{equation}
\begin{aligned}
\frac{\gamma}{\delta} &>> \frac{v_{th}}{v(r)} &&\rm{for} \  \mu = 1. \\[1em]
\end{aligned}
\label{SCE}
\end{equation}

Equation~\ref{SCE} states that the Sobolev approximation is most applicable when the density field approaches a constant and the velocity field gradient is large relative to the Doppler width.  When this is the case, one can factor the density out of the integral in the expression for $\tau_R$, and $\phi_{\nu}$ approaches a delta function in position space, $r$.  Under these conditions, the optical depth will converge to the Sobolev optical depth, $\tau_S$, or the expression defined by Equation~\ref{tau_s}.  We test this claim in the left panel of Figure~\ref{f13} by showing the normalized difference between $\tau_R$ and $\tau_S$ as a function of $\delta$ for various values of $\gamma$.  The X's mark the location of a constant mass outflow rate for each $\gamma$.  From the plot, it is clear that $\tau_S$ is a good approximation for $\tau_R$ in the case of a constant mass outflow rate or when the density field decays slower than is necessary to maintain a constant mass outflow rate  (i.e., for $\delta < \gamma +2$).  For density fields which decay quicker than is necessary to maintain a constant mass outflow rate (i.e., $\delta > \gamma +2$) the values of $\tau_R$ and $\tau_S$ start to diverge and the rate of divergence with $\delta$ grows more slowly with larger $\gamma$ or steeper velocity fields.  Therefore, for the range $0.5 \leq \gamma \leq 2.0$, the Sobolev optical depth is a good approximation for the true optical depth computed by RASCAS which explains the good agreement between the predictions of RASCAS and SALT in Figure~\ref{f11}.   

While Equation~\ref{SCE} can explain the agreement between RASCAS and SALT in Figure~\ref{f11} well, it does not explain the divergence between the predictions of RASCAS and SALT in Figure~\ref{f10}.  In particular, it does not explain the appearance of the redshifted absorption in the RASCAS spectrum which increases in equivalent width with increasing density.  The reason for this shortcoming lies in the fact that the analysis surrounding Equation~\ref{SCE} assumes that there is an infinite amount of material to compute the optical depth.  In reality, however, the outflow has a well defined boundary.  Indeed, SALT will automatically set $\tau_S = 0$ for all photons which are not at resonance with the outflow, while RASCAS will always compute a value for $\tau_R$ which grows in magnitude based on the density and Doppler shift relative to points in the flow.  This explains the discrepancy between the predictions of SALT and RASCAS in Figure~\ref{f10}.  We further verify this claim by examining how the profile changes with $v_{th}$ in the right panel of Figure~\ref{f13}.  We see that the equivalent width of the redshifted absorption features increase when increasing the value of $v_{th}$.  This is because by increasing $v_{th}$, we are increasing the width of the profile function $\phi_{\nu}$, and increasing the amount of photons which can be absorbed by the outflow according to RASCAS, but are otherwise neglected by SALT.  Since this discrepancy between RASCAS and SALT will always be present regardless of the SALT parameters chosen, it is fair to ask if we can still recover the bulk properties of flows when assuming the Sobolev approximation.  We investigate this question in the next section, when we test our ability to recover the SALT parameters from mock spectra generated by RASCAS assuming different models of turbulence.      

\section{Discussion}
\label{Discussion}

The comparison tests between SALT and the low thermal velocity limit of RASCAS performed in section~\ref{lowthermalvelocitytests} demonstrate that the analytical approximations used in SALT can accurately reproduce the results of the Monte Carlo scattering experiments in RASCAS to within a high level of signal to noise.  While this analysis does provide some reassurance into the SALT formalism, it does not address the question of whether or not we can use the idealized models of SALT to gather physically meaningful information from real data.  Indeed, in section~\ref{sc}, we saw that the predictions of RASCAS and SALT begin to diverge when a turbulent/thermal velocity component is included in RASCAS.  Turbulence is thought to play a key role in the formation of outflows \citep{Elmegreen2004,Scalo2004,Hayward2017} as well as in setting their multiphase structure \citep{Fielding2020}.  As such, turbulence is surely a key physical aspect of reality missing from the SALT model.  Differences between the predicted spectra are limited to a few isolated regions, however, and it is worth asking if there is still enough information contained in the spectrum for SALT to recover the bulk properties of the flow.  We begin to answer these questions in this section by testing our ability to recover parameters from mock spectra of idealized outflows which include a turbulent/thermal velocity component.

\subsection{Zero Thermal Velocity Limit}

To begin, we first consider how well SALT can recover parameters from mock spectra generated directly from the idealized models of SALT (i.e., we assume $v_{th} = 0 \ \rm{km} \ \rm{s}^{-1}$).  These tests will act as a control and allow us to gauge the success of our fitting procedure as well as test for any model degeneracy in the SALT parameter space.  To perform the model fits, we use python's emcee package \citep{Foreman-Mackey2013}, which relies on an implementation of Goodman’s and Weare’s Affine Invariant Markov Chain Monte Carlo (MCMC) Ensemble sampler \citep{Goodman2010}.  Similar tests were performed by \cite{Carr2018}, but we repeat the tests here because SALT has been modified since then and includes additional parameters. 

We generated 100 mock spectra with SALT by uniformly sampling parameters from the following ranges. $30^{\circ}\leq \alpha \leq 90^{\circ}$, $0^{\circ}\leq \psi \leq 90^{\circ}$, $2 \ \rm{km} \ \rm{s}^{-1} \leq v_0 \leq 100 \ \rm{km} \ \rm{s}^{-1}$, $100 \ \rm{km} \ \rm{s}^{-1} \leq v_{\infty} \leq 1250 \ \rm{km} \ \rm{s}^{-1}$, $-2 \leq \log{(\tau \ [\rm{\AA}^{-1}])} \leq 3 $, and $0.5 \leq \gamma \leq 2$. The power law index of the density field, $\delta$, was chosen to allow for deviations from a constant mass outflow rate by setting $\delta = \gamma + 2.0 + \Delta\delta$ where $\Delta \delta$ was drawn from a flat distribution ranging from -1.5 to 1.5.  We assumed the wind was fully captured by the aperture (i.e., we set $v_{ap} = v_{\infty}$).  We did not account for holes or clumps in the outflow during this work, and have chosen to set $f_c = 1$ (see \citealt{Carr2018}).  In addition, we have chosen to set the dust opacity, $\kappa = 0$, assuming dust extinction can be constrained by other means.  We assumed Normal measurement errors with variance consistent with observed measurements in COS observations. We simulated the observed spectra by drawing each spectral element from a normal distribution centered on the predicted value and set the standard deviation equal to the error.  The data was smoothed to a resolution of $20 \ \rm{km} \ \rm{s}^{-1}$ and binned to achieve 3 pixels per resolution element.  We explore the effects of resolution on our ability to recover parameters by performing the same experiments at a resolution of $100 \ \rm{km} \ \rm{s}^{-1}$, or the approximate resolution of the NIRSpec G235H mode of JWST, in the appendix.

\begin{figure*}
  \centering
\includegraphics[scale=0.55]{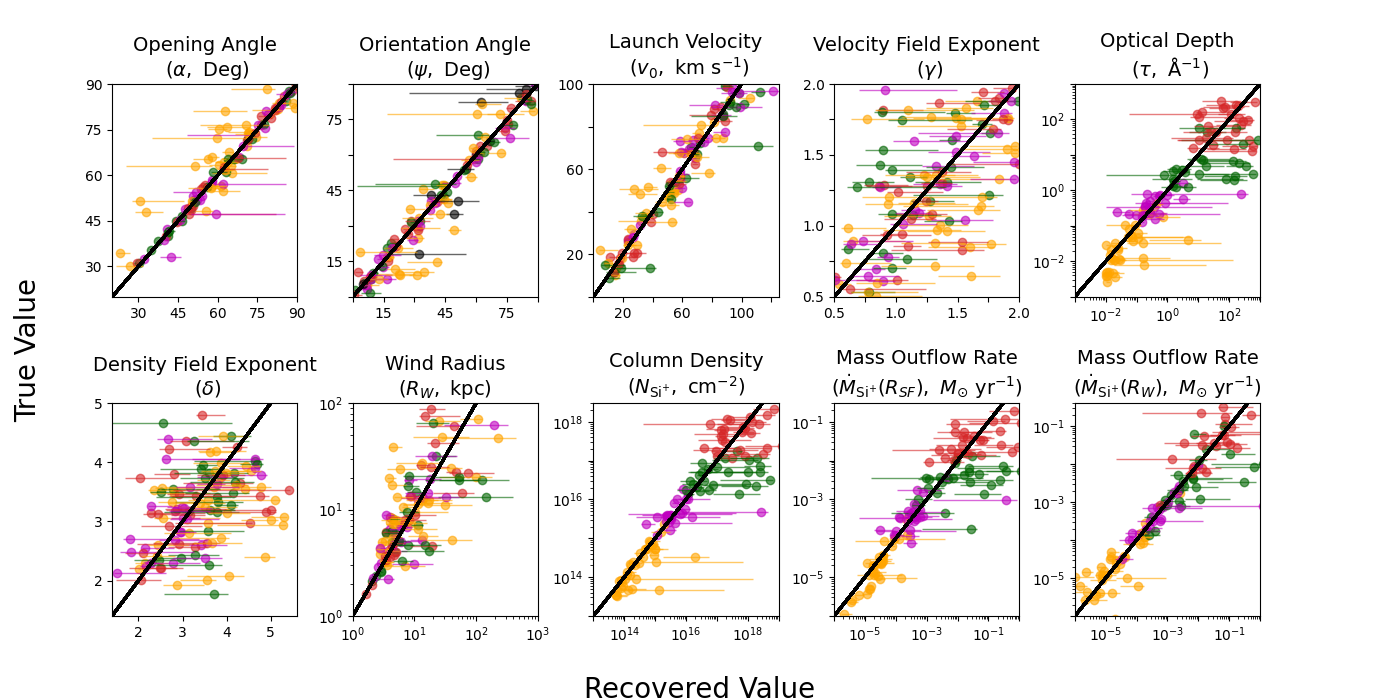}
 \caption{Recovered (horizontal axis) vs true values (vertical axis).  Scatter plots of 100 model fits to mock spectra generated with SALT (i.e., assuming $v_{th} = 0$).  Each spectrum has a resolution of $20 \ \rm{km} \ \rm{s}^{-1}$ and a signal to noise of $10$.  Perfect recovery is indicated by the black lines. All points in the second panel with $\alpha > 85^{\circ}$ have been highlighted in black.  The orientation angle $\psi$ is undefined for spherical flows, and it is unclear what the best fit represents for these points.}
 \label{f14}
\end{figure*}

We assume uniform priors over identical ranges throughout the fitting procedure.  In practice, $v_{\infty}$, $\kappa$, and $v_{ap}$ can be constrained directly from data ($v_{\infty}$ from the absorption limit, $\kappa$ from more traditional means such as Balmer ratios, and $v_{ap}$ by identifying the size of the star forming region from UV images [see \citealt{Carr2021}]).  Thus, we have chosen not to fit to these parameters.  Scatter plots of the recovered parameters (horizontal axis) vs the true values (vertical axis) are shown in the panels of Figure~\ref{f14} where the best fit (i.e., recovered) parameters were chosen to maximize a Gaussian likelihood function (see \citealt{Carr2021}).  The error bars on each point were drawn from the marginal PDFs recovered by the Monte Carlo fitting procedure.  Specifically, the lower (upper) error bar represents the median value of the deviation from below (above) the best fit parameter.  

We include the wind radius, $R_W$, column density of $Si^{+}$, $N_{Si^{+}}$, and mass outflow rate of $Si^{+}$, $\dot{M(R_i)}_{Si^{+}}$, at $R_{SF}$ and $R_W$ in the bottom right panels of Figure~\ref{f14}.  Each of these quantities depends on multiple SALT parameters.  The wind radius is computed as 
\begin{eqnarray}
R_W = R_{SF}\left(\frac{v_{\infty}}{v_{0}}\right)^{1/\gamma},
\end{eqnarray}
the column density as
\begin{gather} 
N_{Si^{+}} =
\begin{cases}
\frac{n_{0,Si^{+}} R_{SF}}{\gamma}\ln{\left(\frac{v_{\infty}}{v_0}\right)}&\ \rm{if} \  \delta = 1 \\[1em]
\frac{n_{0,Si^{+}} R_{SF}}{1-\delta}\left[ \left(\frac{v_{\infty}}{v_0}\right)^{\frac{1-\delta}{\gamma}} - 1\right] & \ \rm{otherwise,}\\
\end{cases}
\label{SF_abs2}
\end{gather}
and the mass outflow rate of $Si^{+}$ as
\begin{eqnarray}
\dot{M}_{Si^{+}}(v) &=& 4\pi(1-\cos{\alpha)}R_{SF}^2 \nonumber\\
&& \times m_{Si^{+}}n_{0,Si^{+}}v_0 \left( \frac{v}{v_0} \right)^{\frac{2+\gamma - \delta}{\gamma}},
\label{MOR2}
\end{eqnarray}
where we have written Equations~\ref{SF_abs2} and \ref{MOR2} in terms of the velocity instead of radius using Equation~\ref{Velocity_Equation}.  $m_{Si^{+}} = 4.66 \times 10^{-23} \ \rm{g}$ is the mass of silicon and $n_{0,Si^{+}}$ is the number density of $Si^{+}$ at $R_{SF}$.  We set $R_{SF} = 1 \ \rm{kpc}$ in all models.  We have color coded all points according to the following density ranges in units of $\rm{cm}^{-3}$. orange: $10^{-7} \leq n_{0,Si^{+}} < 10^{-6}$, magenta: $10^{-6} \leq n_{0,Si^{+}} < 10^{-5}$, green: $10^{-5} \leq n_{0,Si^{+}} < 10^{-4}$, and red: $10^{-4} \leq n_{0,Si^{+}} \leq 10^{-3}$.  

Overall, we found that all parameters are well recovered, except for the power law exponents of the velocity ($\gamma$) and density ($\delta$) fields.  By inspecting Figure~\ref{f11}, one can see that the spectral predictions of SALT for different values of $\gamma$ and $\delta$ are very slight which can explain the degeneracy.  After rerunning these tests at a $S/N = 20$, we found that we are able to better recover these quantities.  Achieving this level of signal to noise should become more practical with the enhanced gathering power of future telescopes such as the 6m-class UV space telescope prioritized by the \cite{Astro2022}.  

Interestingly, the derived quantities such as the mass outflow rate and column density were among the best recovered as they show a small degree of scatter from the perfect 1:1 relation.  The wind radius does appear to be less well recovered compared to the mass outflow rates and column density.  This fact likely reflects the weaker dependence the mass outflow rates and column density have on $\gamma$ and $\delta$ compared to the wind radius.  Indeed, the exponents in Equations~\ref{SF_abs2} and \ref{MOR2} are normalized by $\gamma$.  

The degree of scatter starts to increase with larger values in all quantities proportional to density (i.e., the red points).  This reflects the inherent degeneracy in saturated lines - that is, once all of the photons have been absorbed by the outflow at a given wavelength, one cannot detect the addition of more material at resonance with this wavelength.  Lastly, our ability to recover geometric information such as the opening angle, $\alpha$, diminishes with decreasing density (i.e., the orange points).  This reflects the weakening of absorption and emission features relative to the noise with decreasing density.  We found that $\alpha$ was better recovered during the $S/N = 20$ tests which supports this claim.

\begin{figure*}
  \centering
\includegraphics[scale=0.55]{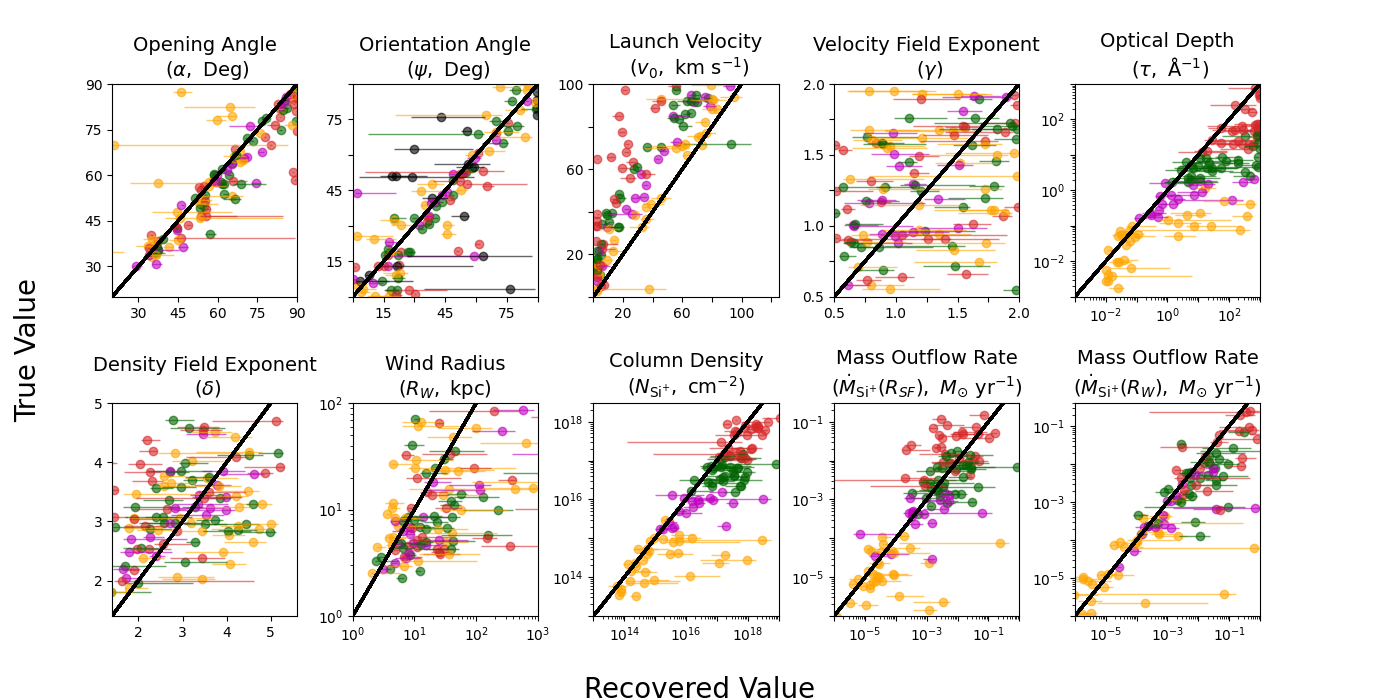}
 \caption{Same setup as Figure~\ref{f14}, except each mock spectrum was generated with RASCAS to include a uniform turbulence model with Doppler parameter fixed at $10 \ \rm{km} \ \rm{s}^{-1}$.}
 \label{f15}
\end{figure*} 

\begin{figure*}
  \centering
\includegraphics[scale=0.55]{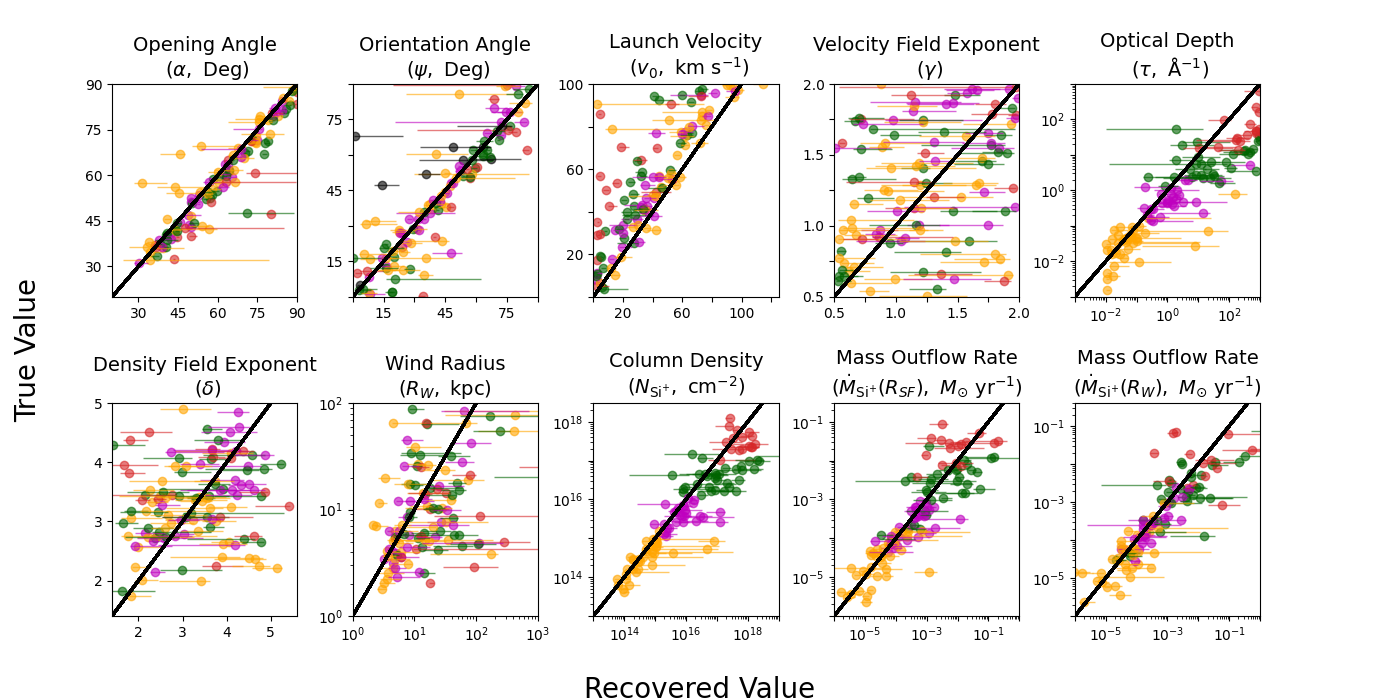}
 \caption{Same setup as Figure~\ref{f14}, except each mock spectrum was generated with RASCAS to include a radially dependent turbulence model with Doppler parameter which scales as $10\%$ of the velocity field (i.e., $v_{th} = 0.1v(r)$).}
 \label{f16}
\end{figure*}

\subsection{Turbulent/Thermal Line Broadening}

Here we repeat the procedure above except this time we generate mock spectra using RASCAS and include the effects of turbulent/thermal motion set by the Doppler parameter, $v_{th}$.  We consider two basic models for turbulence: a uniform model with constant speed throughout the outflow and a radially dependent model which scales as $10\%$ of the flow speed (i.e., $v_{th} = 0.1v(r)$).  The former is based off of the high resolution Cholla Galactic OutfLow Simulations (CGOLS) suite by \cite{Schneider2018a} which measure turbulence to gradually change with scale height from roughly $25 \ \rm{km} \ \rm{s}^{-1}$ to $10 \ \rm{km} \ \rm{s}^{-1}$ starting at the base of the outflow in the CGM.  The latter comes from the analytical models by \cite{Fielding2020} who consider a shear driven turbulence model for the mixing of cool and hot gas in the CGM.  For the uniform model, we considered Doppler parameters equal to $25 \ \rm{km} \ \rm{s}^{-1}$ and $10 \ \rm{km} \ \rm{s}^{-1}$ spanning the range measured by \cite{Schneider2018a}.  

We found that the $25 \ \rm{km} \ \rm{s}^{-1}$ RASCAS model showed exceptionally enhanced fluorescent emission which is not observed in the spectra of local galaxies (e.g., \citealt{Berg2022}). For this reason, we did not include this model in our results.  We suspect that RASCAS' inability to account for holes (i.e., $f_c$) in the outflows is a major reason for the unrealistic features in these mock spectra.  Including holes in the outflows will reduce the number of times a photon can scatter and thereby reduce its probability of getting absorbed in the wings of the cross section.  Dust, which acts to reduce reemission in normalized profiles (see \citealt{Carr2021}), should also play a factor.  The $10 \ \rm{km} \ \rm{s}^{-1}$ model shows spectra more consistent with reality and we have decided to include them in our analysis in Figure~\ref{f15}, but we advise the reader to proceed with caution by viewing the results of these tests more as the results of a mathematical exercise rather than an experiment done with real data.

Overall, we see the same trends as in Figure~\ref{f14}, however, some biases do emerge.  $v_0$ appears to be consistently underestimated by SALT, and this bias is a function of the density.  This can be expected because adding thermal/turbulent motion increases the amount of absorption occurring at velocities less than $v_0$ (the value of $v_0$ can be used roughly to indicate where absorption begins in observed velocity space).  Thus SALT is underestimating $v_0$ to compensate for this extra absorption.  The fact that the error is consistent regardless of the value of $v_0$ verifies our claims in section~\ref{sc} that the differences between RASCAS and SALT (i.e., the redshifted absorption) depend only on $v_{th}$ and the density.  Similarly, SALT can obtain more absorption by increasing the optical depth, $\tau$, as well as by increasing the amount of material available to absorb by increasing the size of the opening angle of the flow, $\alpha$.  

Interestingly, the dependent quantities - wind radius, column density, and mass outflow rate - appear to be well recovered again, albeit the wind radius does appear to be overestimated in several cases and the column density shows a slight bias towards being overestimated.  This suggests that the biases observed in the recovered values for individual parameters cancel out in the calculation of the dependent quantities.               

We consider the radially dependent model, $v_{th} = 0.1v(r)$, in Figure~\ref{f16}.  We have implemented a lower bound of $2.8 \ \rm{km} \ \rm{s}^{-1}$, or about the thermal speed of $Si^{+}$ gas at a temperature of $10^4 \ \rm{K}$, to ensure that we do not obtain an unreasonably small value of the Doppler parameter for all locations in the outflow.  This cutoff is similar to the lower limit studied by \cite{Fielding2022} who assumed an injection velocity of $30 \ \rm{km} \ \rm{s}^{-1}$.  Overall, we see similar patterns emerge as in the uniform turbulence model, but the results appear to be better recovered.  This is because, by design, this model achieves weaker turbulent speeds at the densest portion of the outflow near the launch radius.  Thus, in the regions where most of the scattering is taking place, photons have a smaller chance of getting absorbed in the tails of the cross section.

\section{Conclusions}
In this paper, we tested the semi-analytical line transfer (SALT) model for predicting galactic spectra against the numerical radiation transfer code, RASCAS, in the context of idealized models of biconical outflows.  Our comparisons between SALT and RASCAS inspired upgrades to both models which we presented here for the first time.  These include an alternative technique for computing the SALT model which considers the absorption of non-radially traveling photons by an expanding envelope and a customizable adaptive mesh refinement (AMR) routine for RASCAS capable of resolving the features of galactic winds including edges or boundaries, the velocity and density field gradients, and the optical depth. 

We tested how well SALT can recover the bulk properties of flows from mock spectra in the cases of non-turbulent and turbulent gases.  In regards to the former, after invoking observable constraints on the aperture, dust opacity, and terminal velocity, we found that SALT can recover the geometry (opening angle and orientation) of the bicones, the optical depth, column density, mass outflow rate, and wind radius well, but struggled to constrain the velocity and density field gradients when simulated data with $S/N = 10$ and $R = 20 \ \rm{km} \ \rm{s}^{-1}$ are used.  We found that by increasing the signal to noise ratio one can start to better constrain the gradients.  In regards to the latter, at the same resolution and $S/N$, we found that biases appear in the recovery of the independent parameters listed above, but found that the biases tend to cancel out in the calculation of the dependent quantities such as the column density, wind radius, and mass outflow rate.  We reran these tests at a lower resolution of $R = 100 \ \rm{km} \ \rm{s}^{-1}$ and obtained similar results, albeit with a larger degree of scatter in the recovered values.  

The results of our comparisons confirm the accuracy of the SALT model with numerical calculations, and we deem it a suitable model for interpreting future data sets at high ($R \sim 20 \ \rm{km} \ \rm{s}^{-1}$) and medium ($R \sim 100 \ \rm{km} \ \rm{s}^{-1}$) resolution at a moderate signal to noise of 10.  If the idealized model configurations can accurately describe galactic environments, then the SALT model should be particularly effective at constraining metal outflow rates and the column densities of metal ions.  

\begin{acknowledgments}
C.C. would like to thank the Centre de Recherche Astrophysique de Lyon (CRAL) for hosting him during a portion of his time spent on this research.  He would also like to thank the faculty at CRAL for providing excellent recommendations for experiencing (high and low) French culture, and especially, for introducing him to choux chantilly.  We would like to thank Thibault Garel who developed and provided us with the first version of the idealised models on regular grid in RASCAS. We also thank Thibault Garel for insightful discussions.  We acknowledge the Minnesota Supercomputing Institute (MSI) at the University of Minnesota for providing the majority of computational support for this project (URL: http://www.msi.umn.edu).  Some simulations were computed at the Common Computing Facility (CCF) of the LABEX Lyon Institute of Origins (ANR-10-LABX-0066). We also acknowledge the Pôle Scientifique de Modélisation Numérique (PSMN) of the ENS de Lyon for the computing resources.  Support for program number HST-AR-16606 was provided by NASA through a grant from the Space Telescope Science Institute, which is operated by the Association of Universities for Research in Astronomy, Incorporated, under NASA contract NAS5-26555.  
\end{acknowledgments}

\appendix

\section{Low Resolution Results}

\begin{figure*}
  \centering
\includegraphics[scale=0.55]{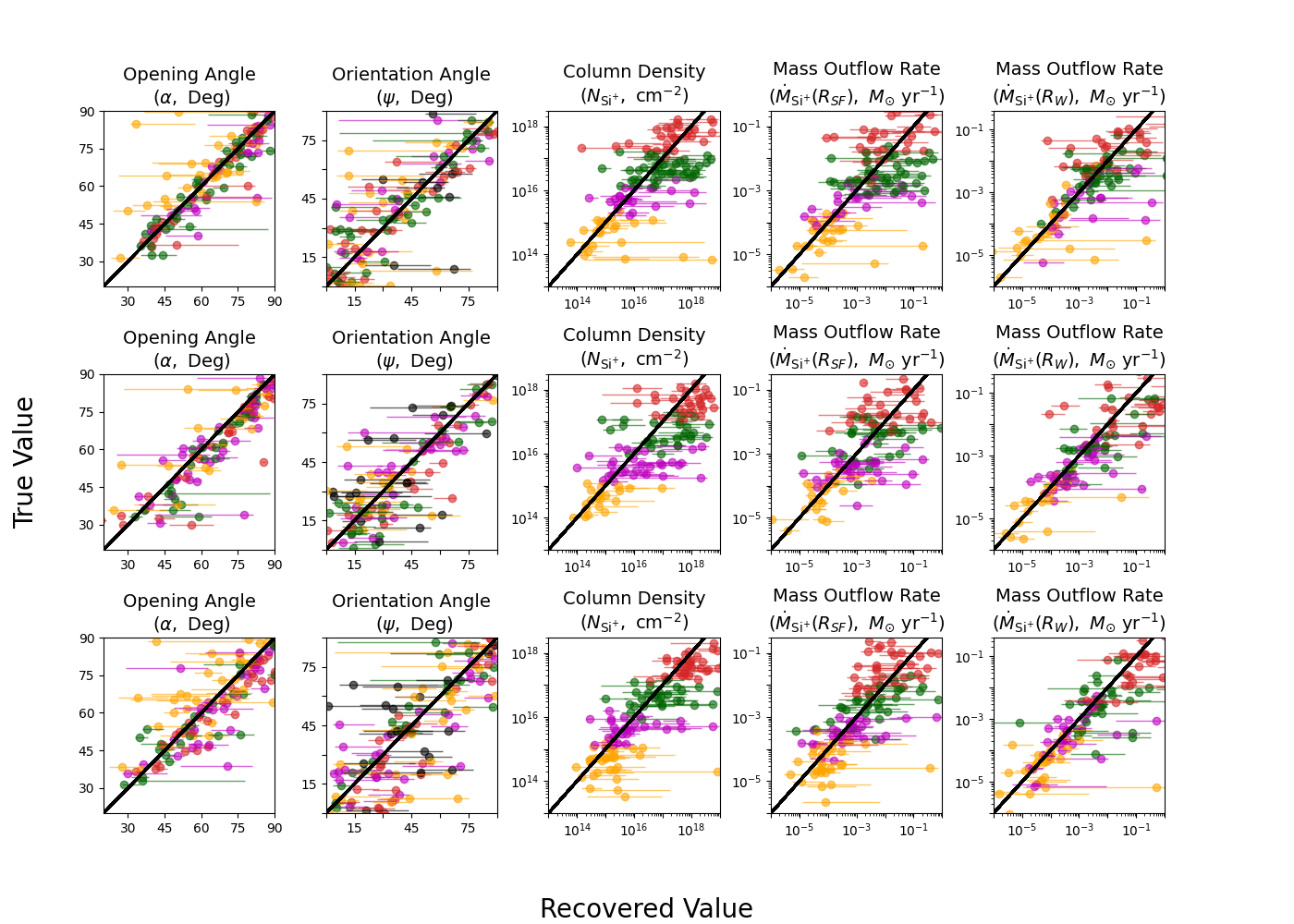}
 \caption{Same setup as Figure~\ref{f14}, except each spectrum has been smoothed to a resolution of $100 \ \rm{km} \ \rm{s}^{-1}$. \textbf{\emph{Top Row}} Mock spectra were generated with SALT with Doppler parameter $v_{th} = 0 \ \rm{km} \ \rm{s}^{-1}$.  \textbf{\emph{Middle Row}} Mock spectra were generated with RASCAS with Doppler parameter $v_{th} = 10 \ \rm{km} \ \rm{s}^{-1}$.  \textbf{\emph{Bottom Row}} Mock spectra were generated with RASCAS with Doppler parameter $v_{th} = 0.1v(r) \ \rm{km} \ \rm{s}^{-1}$.}
 \label{f17}
\end{figure*}

Here we perform the same tests as in section~\ref{Discussion}, but at a resolution of $100 \ \rm{km} \ \rm{s}^{-1}$, or the approximate resolution of the NIRSpec G235H mode of JWST.  We show the results for all three different turbulence models: $v_{th} = 0, 10, \ \rm{and} \ 0.1v(r) \ \rm{km} \ \rm{s}^{-1}$ in the the top, middle, and bottom rows of Figure~\ref{f17}, respectively.  Because decreasing the resolution will only lesson our ability to recover parameters, we only show the outflow rates, column density, opening angle ($\alpha$), and orientation angle ($\psi$).  The latter three quantities can be used to infer the escape fraction of ionizing radiation.  All of the displayed quantities are well recovered, albeit with a larger degree of scatter compared to the $R = 20 \ \rm{km} \ \rm{s}^{-1}$ results, regardless of the turbulence model assumed.  This result is promising because it suggests we can use models such as SALT to interpret JWST data to constrain outflow rates and the escape fraction of ionizing radiation in the early universe.

\section{SALT parameters}
In Table~\ref{tab:pars} we report the primary SALT parameters with their symbol and their definition. The second half of the table defines additional support parameters used in the calculation of  models in Section~2. 

\begin{table*}[t]
	\center
	\caption{List of SALT parameters and definitions. }
	\begin{tabular}{cccc}\hline\hline
		Symbol&Definition&Symbol&Definition\\
		\hline 
		$R_{SF}$ & Radius of continuum source& $R_W$ & Wind radius\\
		$\alpha$&Half opening angle &$\psi$ &Orientation angle \\ 
		$v_0$ &Launch velocity & $v_{\infty}$ &Terminal velocity\\
		$R_{AP}$ & Aperture radius & $v_{ap}$ &Velocity field at $R_{AP}$ \\
		$\gamma$ &Velocity field power law index &$\delta$ &Density field power law index \\
		$\tau_0$ &Optical depth & $\tau$ & $\tau_0$ divided by $f_{ul} \lambda_{ul}$\\
		$\kappa$ &Dust opacity multiplied by $R_{SF} n_{\rm{dust},0}$&$f_c$ &Covering fraction inside the bicone\\
		$n_0$ & Number density at $R_{SF}$\\
		\hline
		$y$& Shell velocity normalized by $v_0$&$y_\infty$&Terminal velocity normalized by $v_0$\\
		$\tau_S$&Sobolev optical depth&$\phi$ & Angle between the velocity and photon ray\\
	$\Omega_x$&Surface of constant observed velocity $x$ &$\Gamma_x$ & Intersection of $\Omega_x$ with the $s\xi$-plane\\
	$\Xi$& $\xi$ coordinate of $\Gamma_x$&$S$& $s$ coordinate of $\Gamma_x$\\
		$f_{lu}$&Oscillator strength for the $lu$ transition&$\lambda_{lu}$&Wavelength for the $lu$ transition\\
			$\Theta_C$& Max angle for continuum absorption&$f_g$ & Geometric scale factor for wind geometry\\
		\hline 
		\vspace{0.4cm}
	\end{tabular}
	\label{tab:pars}
\end{table*}  

\begin{table*}[t]
\caption{Atomic Data for Si II ion.  Data taken from the NIST Atomic Spectra Database$^{\MakeLowercase{a}}$.}
\begin{tabular}{P{.7cm}|P{2.3cm}|P{1.4cm}|P{1.6cm}| P{2.8cm}|P{.9cm}|P{2.6cm}|P{2.8cm}}\hline\hline
Ion & Vac. Wavelength&  $A_{ul}$&$f_{ul}$& $E_{l}-E_{u}$&$g_l-g_u$&Lower Level&Upper Level\\
&\AA& $s^{-1}$ &&$eV$&&Conf.,Term,J&Conf.,Term,J\\[.5 ex]
\hline 
 Si II& $1190.42$ &$6.53\times 10^{8}$&$2.77\times 10^{-1}$&$0.0000-10.41520 $&$2-4$ &$3s^23p, {}^2P^0, 1/2$&$3s3p^2, {}^2P, 3/2$ \\
&$1193.28$ & $2.69\times 10^{9}$ &$5.75\times 10^{-1}$&$0.0000-10.39012 $&$2-2$&$3s^23p, {}^2P^0, 1/2$&$3s3p^2, {}^2P, 1/2$  \\
& $1194.50$ & $3.45\times 10^{9}$ &$7.37\times 10^{-1}$&$0.035613-10.41520$&$4-4$&$3s^23p, {}^2P^0, 3/2$&$3s3p^2, {}^2P, 3/2$ \\
& $1197.39$ & $1.4\times 10^{9}$ &$1.5\times 10^{-1}$&$0.035613-10.39012$&$4-2$&$3s^23p, {}^2P^0, 3/2$&$3s3p^2, {}^2P, 1/2$ \\[1ex]
\hline
\end{tabular}
\label{tab:atomicdata}
\\$^{\text{a}}$ http://www.nist.gov/pml/data/asd.cfm
\end{table*}

\bibliographystyle{jphysiscsB} 
\bibliographystyle{apj} 
\bibliography{doc}

\end{document}